\documentclass[twocolumn,floatfix,superscriptaddress,amsmath,showpacs,aps,prb]{revtex4-1}

\usepackage{t1enc}
\usepackage[final]{graphicx}
\usepackage{epsfig}
\usepackage{bm}
\usepackage[normalem]{ulem}
\usepackage[usenames]{xcolor}
\usepackage{times}
\usepackage{verbatim}
\usepackage{amssymb}

\newcommand{\beq}{\begin{equation}}
\newcommand{\eeq}{\end{equation}}
\newcommand{\beqa}{\begin{eqnarray}}
\newcommand{\eeqa}{\end{eqnarray}}
\newcommand{\beqann}{\begin{eqnarray*}}
\newcommand{\eeqann}{\end{eqnarray*}}

\begin{document}

\title{Quantum and thermal melting of stripe forming systems with competing long ranged interactions}
\author{Alejandro Mendoza-Coto}
\email{alejandro.mendoza@ufsc.br}
\affiliation{Departamento de F\'\i sica, Universidade Federal de Santa Catarina, 88040-900, Florian\'opolis, SC, Brazil}
\author{Daniel G. Barci}
\affiliation{Departamento de F{\'\i}sica Te\'orica,
Universidade do Estado do Rio de Janeiro, Rua S\~ao Francisco Xavier 524, 20550-013,  
Rio de Janeiro, RJ, Brazil}
\author{Daniel A. Stariolo}
\email{stariolo@if.uff.br}
\affiliation{Departamento de F\'{\i}sica,
Universidade Federal Fluminense and
National Institute of Science and Technology for Complex Systems\\
Av. Gal. Milton Tavares de Souza, s/n, Campus da Praia Vermelha, 24210-346,
Niter\'oi, RJ, Brazil}

\date{\today}

\begin{abstract}
We study the quantum melting of stripe phases in models with competing short range and long range interactions decaying
with distance as $1/r^{\sigma}$ in two space dimensions. At zero temperature we find a two step disordering of the stripe phases with the growth
of quantum fluctuations. A quantum critical point separating a phase with long range positional order from a phase with long
range orientational order is found when $\sigma \leq 4/3$, which includes the Coulomb interaction case $\sigma=1$. For $\sigma > 4/3$ the
transition is first order, which includes the dipolar case $\sigma=3$. Another quantum critical point separates the orientationally
ordered (nematic) phase from a quantum disordered phase for any value of $\sigma$. Critical exponents as a function of $\sigma$ are computed
at one loop order in an $\epsilon$ expansion and, whenever available, compared with known results. For finite temperatures it is found
that for $\sigma \geq 2$ orientational order decays algebraically with distance until a critical Kosterlitz-Thouless line. Nevertheless,
for $\sigma < 2$ it is found that long range orientational order can exist at finite temperatures until a critical line which terminates
at the quantum critical point at $T=0$. The temperature dependence of the critical line near the quantum critical point is determined as
a function of $\sigma$.
\end{abstract}





\maketitle

\section{Introduction}
\label{Sec:Introduction}
Quantum systems with fluctuating stripe order have received an increasing attention of the condensed matter community
in the last years\cite{KivFrad2003,Vojta2009,FrKiTr2015}.  Different quasi-two-dimensional systems such as quantum Hall systems\cite{Fogler1996,Lilly1999,FrKi1999,Friess2014}, 
cuprates and iron-based  superconductors\cite{Parker2010,Daou2010,Tanatar2016} as well as heavy fermion compounds\cite{Grigera2001,Borzi2007} present inhomogeneous and/or anisotropic electronic/spin structures  at very low temperatures.  
A common feature of these systems is the possible  existence of a quantum critical point (QCP) separating phases that break  translational and/or rotational symmetries\cite{Taillefer2010}.  

With growing temperature stripe patterns melt, typically by the proliferation of dislocations and disclinations. The theory of thermal melting of stripes is reasonably well understood\cite{ToNe1981,AbKaPoSa1995,Wexler2001}. When quantum fluctuations play a role, as in the electronic systems cited above,  the phenomenology of anisotropic phases may be much
more complex than in the corresponding thermal, or classical counterpart. On general grounds it is expected that upon lowering the temperature a crossover must occur
between thermal and quantum dominated fluctuations. Eventually, quantum fluctuations may be responsible for a true $T=0$ quantum critical point (QCP). The behavior of
thermodynamic and electronic transport properties must be strongly influenced by the proximity of the QCP, and then it is important to characterize the different possible
universality classes of quantum phase transitions\cite{Sa1997,Mucio2001,Sa2011}. A quantum theory of stripe melting which can account for this rich
phenomenology is still under development.  

A model for continuous stripe quantum melting in metals, driven by a particular type of topological defects, double dislocations,  was recently proposed in 
connection with the physics of cuprate superconductors~\cite{Senthil2012}. This model considers the simultaneous presence of charge and spin stripes. As a
consequence, it is argued that single charge stripe dislocations (which would lead to frustration in the spin stripes) remain gapped, justifying that the physics of charge stripe
melting should be mainly governed by the proliferation of double dislocations. Upon disordering, this leads to a ``stripe loop metal'' phase, different from the
usual Fermi liquid phase. When spin order is not considered, the model leads to similar results as we reach in the
present study, which only deals with one type of degrees of freedom.

In the context of quantum Hall phases, the condensation of lattice defects  was recently studied\cite{Cho2015}. A very general framework of a gauge field theory of quantum liquid crystals at $T=0$ in continuum space has also been proposed and recently reviewed in [\onlinecite{Beekman2016}].

Another context in which the results of the present work can be relevant is the field atomic gases\cite{Radzihovsky-2012} and in particular in ultra-cold dipolar Fermi gases~\cite{Lu2012,Park2015}. 
The experimental realization of ultra-cold atom systems is
progressing rapidly and promises to be an ideal testing ground for the physics of quantum strongly correlated systems. Several electronic liquid
crystalline phases are predicted to be present in these systems~\cite{Yamaguchi2010,Gorshkov2011,Babadi2011,Parish2012}.  Recently the thermal melting of two dimensional dipolar
Fermi gases was addressed in [\onlinecite{Wu2016}]. When all the dipoles are at a generic tilting angle $\Theta$ with respect to the $xy$ plane of the system, the thermal
melting of stripes is found to be driven by the proliferation of dislocations, leading to an effective anisotropic $XY$ model, which then allows to predict the elastic and critical properties straightforwardly. Nevertheless, for tilting angle $\Theta=0$, i.e. when all dipoles are oriented perpendicular to the $xy$ plane, the system recovers rotation
invariance on the plane, and then long range stripe order is forbidden at finite $T$~\cite{Wexler2001,BaSt2009,MeStNi2015,MeStNi2016,NiMeSt2016}. While longe range positional order is forbidden, orientational
order of stripes survives to dislocations proliferation, leading to a nematic-like phase~\cite{Wu2016}. In the present work we show that, for dipolar interactions in a rotationally symmetric system, only quasi-long-range nematic order in the orientation of stripes is possible at finite temperatures.

It is worth mentioning that while most properties of a quantum phase transition are dictated by symmetry, dynamical properties depend essentially on microscopy. In the context of Fermi liquids, Landau damping could produce over-damped modes with dynamical exponent $z=3$\cite{Lawler2006} near the isotropic-nematic phase transition. In the presence of a tetragonal lattice,  Ising-nematic fluctuations could renormalize the dynamical exponent as well\cite{Metlitski2010}, which adds to the complexity of the observed
phenomenology.  

The emergent phenomenology related with striped phases is supposed to come from the  competition between short range attractive 
interactions and long range repulsive ones\cite{EmKi1993,SeAn1995,Chayes1996} at a microscopic level. In this perspective, the
relevance  of  residual long range interactions is an interesting and important problem and, to the best of our knowledge, an open one.   
For instance, in  quantum Hall systems, it is known that long-range Coulomb interactions change the nature of the smectic 
phase\cite{BaFrKiOg2002}.  Moreover, in the context  of charged cold atoms, Coulomb as well as dipole interactions play 
a relevant role\cite{BrNe2014,Wu2016}.  Recently, the effects of long 
ranged interactions in classical two-dimensional stripe melting was considered~\cite{MeStNi2015,MeStNi2016,NiMeSt2016}.  It was shown that, consistent with common assumptions, for  sufficiently
short-ranged interactions, positional correlations are short ranged while a Kosterlitz-Thouless transition from
an  isotropic  to  a  quasi-long-range  nematic  phase
takes place. Interestingly, it was found that for sufficiently long-ranged repulsive interactions a second order phase transition occurs between a disordered isotropic phase and  a truly long-range orientationally ordered nematic phase. This result implies that, for classical models, the lower critical dimension depends on the range of the interaction: the larger the range of the repulsive interaction, the smaller the lower critical dimension. 

In this work we extend the approach to orientational phases in systems with long ranged competing interactions to the realm of quantum melting of stripe patterns. The 
starting point is a coarse-grain effective Hamiltonian for isotropic, competing short range and long range interactions which decay with distance as $1/r^{\sigma}$. This
action is a well known continuum limit of a large family of truly microscopic Hamiltonians, and in this sense the connection of our results with specific microscopic models
is straightforward. Then, starting from the well known stripe ground states of the effective model, we promote the displacement field of the stripe modulation to a quantum time dependent operator,
introduce dislocations and
obtain the relevant dispersion relations for the quantum elasticity theory with long range competing interactions. At finite temperatures the mean squared fluctuations of
the displacement field diverge with the size of the system independently of the range of interactions, and then positional order can be strictly short ranged, a result well
known in models with short range interactions. A more interesting
scenario emerges for the orientational degrees of freedom which are stable to Gaussian fluctuations. Then, after introducing dislocations, the relevant topological defects
in stripe forming systems, we consider the interaction between two far apart portions of the striped
pattern through a multipolar expansion of the relevant density (electronic or magnetic density). This leads to a 
model of quantum rotors in the plane of the system with generalized dipolar interactions~\cite{MeStNi2015}. Analysis of this model shows that, when temperature
fluctuations dominate over quantum ones,
the well known Kosterlitz-Thouless type phase transition driven by disclination unbinding is
restricted to sufficiently short ranged competing interactions, $\sigma \geq 2$. If the interactions are long ranged enough, $\sigma <2$, a genuine second order phase transition
to a phase with long range orientational order may exist at finite $T$. 

At $T=0$ the conclusions from the quantum elastic theory are different: for sufficiently weak quantum
fluctuations the mean square fluctuations of the displacement field may be finite, while they diverge for strong enough fluctuations. This indicates the existence of a quantum
phase transition from a smectic-like phase to a nematic-like one at a critical value of the strength of quantum fluctuations. In the context of a McMillan-deGennes theory
this phase transition turns to be of first order for $ \sigma > 4/3$ and second order for sufficiently long range repulsive interactions $\sigma \leq 4/3$. Subsequently, at higher values of the quantum fluctuations, orientational
order also breaks down leading to a fully disordered phase. For short range interactions, $\sigma \geq 2$, the quantum nematic-isotropic transition in $d=2$ is in the universality class of the thermal XY model in $d=3$. For sufficiently long ranged interactions, $\sigma <2$, the critical exponents depend continuously on the interaction range. 
Finally, the critical line emerging from the quantum critical point at the $T=0$ isotropic-nematic transition is obtained for sufficiently low temperatures. No critical
line emerges from the nematic-smectic transition point because positional order is completely absent at finite temperatures.

These results are summarized in figures (\ref{fig:phasediag1}) and (\ref{fig:phasediag2}).  
In figure (\ref{fig:phasediag1}), we show a  qualitative phase diagram for  $\sigma\geq2$ in two-dimensions. At $T=0$, we show two quantum phase transition points.  At $r_{1c}$,  a  first order transition, form a quantum smectic phase (``Long Range Positional order, L.R.P.O) to a quantum nematic phase (``Long Range Orientational Order, L.R.O.O), takes place.  $r_2^c$ is  a  quantum critical point, separating the nematic phase form the quantum disordered phase (Short Range Orientational Order, S.R.O.O).
As we have anticipating,  at finite temperature, there cannot be true long ranged order, then, there is a Kosterlitz-Thouless line separating a disorder region (S.R.O.O) form a quase long range orientational order region (Q.L.R.O.O) at lower temperatures.
The results for strong long range interactions, $\sigma<2$,  are shown in figure (\ref{fig:phasediag2}).  
The main important diference is  that long range interactions  stabilize a true long range orientational order (L.R.O.O) at finite temperatures.  The critical exponents associated with the quantum critical point, $r_2^c$, are summarized in table  \ref{tb:exponents}.

The organization of the paper is as follows: in Section \ref{Sec:generaltheory} we briefly introduce the model for competing interactions at different scales
and proceed in Subsection \ref{Sec:Melting} to a presentation of the quantum elastic theory of stripe melting in a pedagogical way. In Subsection \ref{Sec:rotors}
we introduce the model of quantum rotors in the plane, from which most of the results will be extracted. Then, in Section \ref{Sec:Criticality} we describe the
main results for the isotropic-nematic phase transitions at $T=0$. Two subsections describe, respectively, the behavior for $\sigma \geq 2$ and $\sigma <2$. 
Complementing the discussion of the $T=0$ phase diagram, in Section \ref{Sec:Smectic} the nematic-smectic phase transition is described. In Section \ref{Sec:Thermal}
the results for the finite $T$ phase diagrams are presented for both $\sigma \geq 2$ and $\sigma <2$. Finally, a discussion of the results and some conclusions are
presented in Section \ref{Sec:Conclusions}.

\section{Quantum theory of stripe melting}
\label{Sec:generaltheory}
The simplest model to describe the effect of competing interactions at different scales in two spatial dimensions can be cast in 
the following  coarse-grained Hamiltonian~\cite{MeStNi2015}:
\begin{eqnarray}
\nonumber
\mathcal{H}[\phi(\vec{x})]&=&\frac{1}{2}\int d^2 x\,\left(\vec{\nabla}\phi(\vec{x}) \right)^2
\nonumber \\ \nonumber
&+&\frac{1}{2}\int d^2x\int d^2x'\ \phi(\vec{x}){J}(\vec{x}-\vec{x'})\phi(\vec{x'})\\
&+& \frac{1}{2\beta}\int d^2x\ V(\phi(\vec{x})) \; .
\label{eq:Ham}
\end{eqnarray} 
$\phi(\vec x)$ represents a density field or scalar order parameter and, in the absence of external fields, the Hamiltonian has the Ising symmetry $\phi\to -\phi$.    $J(\vec x-\vec x')$ is a repulsive long ranged isotropic interaction which decays as a power law of the distance in the form  $J(\vec{x})=J/\vert \vec x\vert^\sigma$. Although it is convenient to work with an arbitrary exponent $\sigma$, physically relevant examples are the Coulomb  interaction in charged systems ($\sigma=1$) and the dipolar interaction between out-of-plane magnetic moments ($\sigma=3$) in magnetic films. 
Additionally, entropic contributions generate a local potential $V(\phi)$ which,  for simplicity,  we consider to be a degenerate double well potential of the form
 $V(\phi)=-\frac{a}{2}\phi^2+\frac{b}{4}\phi^4$, with $a>0$ and $b>0$. 
 
Finally,   $\beta =(k_BT)^{-1}$ is the inverse temperature. 

\subsection{Stripe fluctuations and melting}
\label{Sec:Melting}
\begin{figure}[ht]
\includegraphics[scale=0.45]{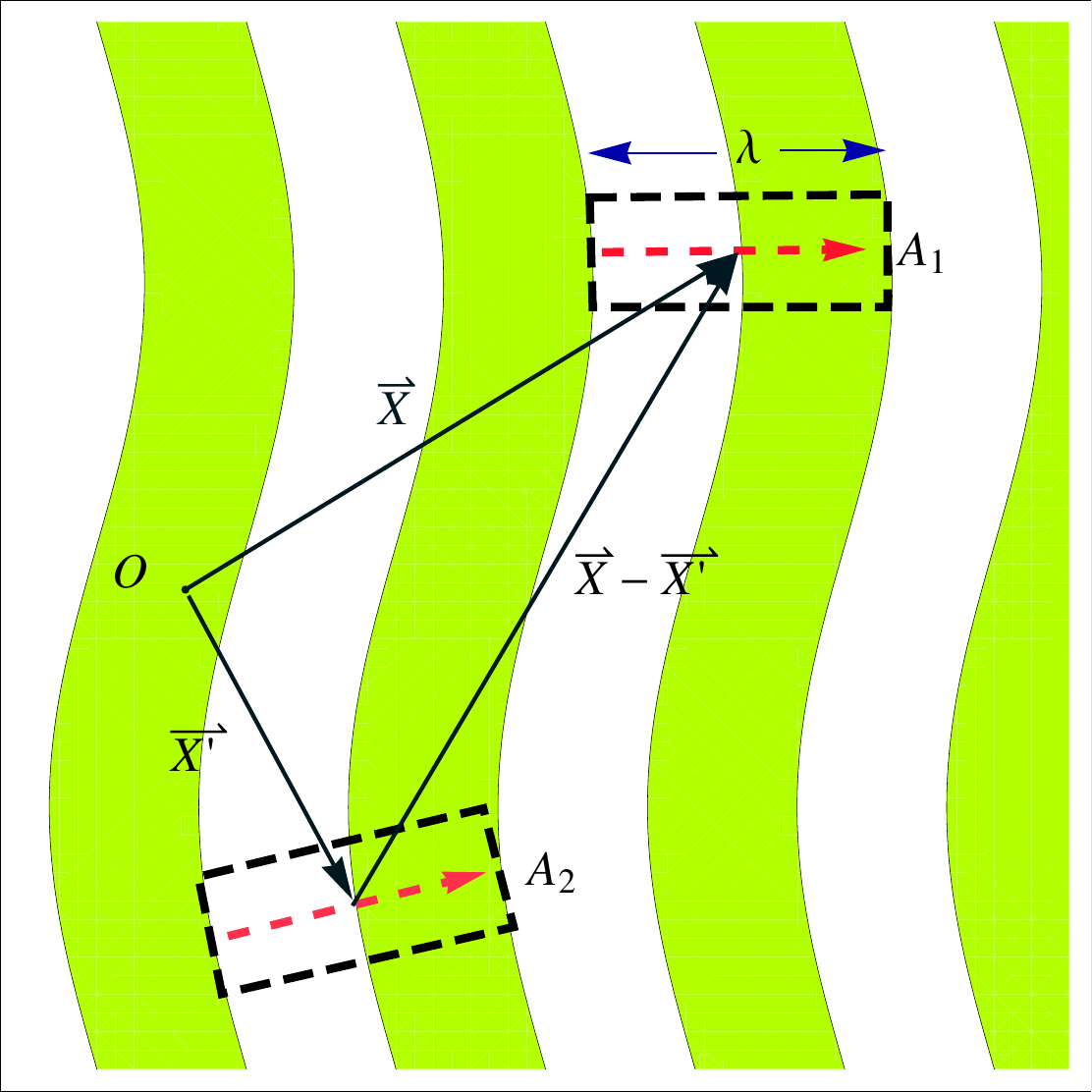}
\caption{Schematic representation a smoothly deformed striped pattern. The red, dashed arrows indicate the orientation of the domain walls and define local dipolar moments. The interaction between two far apart elementary stripe dipoles is shown.}
\label{fig:stripes}
\end{figure}

In the absence of long-range interactions, the system tends to form condensate states, with $\phi=\pm\sqrt{a/b}$.
Due to the Ising symmetry, both condensates are equally probable and the system tends to phase separate.  The long range repulsion $J(|\vec x|)$ frustrates this tendency and the ground state ends being inhomogeneous and/or anisotropic. The simplest and most commonly found configuration of this type is a unidirectional modulation 
characterized by a single wave vector $\vec k_0$~\cite{Br1975,ToNe1981,SeAn1995,BaSt2007}. In this case, the order parameter can be written in the form $\phi(\vec{x})=\sum_n\phi_n \cos(n\vec k_0\cdot  \vec x)$, where the Fourier coefficients $\phi_n$ determine the profile of the modulation. Long wave-length fluctuations can be parametrized by a displacement field $u(x,y)$ in the form 
\begin{equation}
\phi(x,y)=\sum_n\phi_n \cos(n\, k_0[ x+ u(x,y)])\; , 
\label{eq:phiStripe}
\end{equation}
where $x$ is the average direction of the modulation and $k_0$ stands for the modulus of $\vec{k}_0$. The displacement field $u(x,t)$ should be interpreted as the projection of the vector displacement $\vec u$ on the ordering vector $\vec k_0$, {\em i.\ e.\ }, $u=\vec u\cdot \vec k_0$.  A typical configuration is illustrated in Figure \ref{fig:stripes}.  The shaded area represents regions with $\phi(\vec x)>0$ while in the white regions $\phi(\vec x)<0$. 
The Fourier spectrum $\phi_n$ can be computed, in principle, using a mean-field approximation.

Replacing  (\ref{eq:phiStripe}) into  Eq.  (\ref{eq:Ham}), we can compute an effective Hamiltonian in terms of the displacement field $u$. Making a long wave-length approximation (a gradient expansion) and keeping the leading quadratic terms we obtain the elastic energy~\cite{MeStNi2015}
\begin{equation}
\mathcal{H}=\mathcal{H}_0+
\frac{1}{2}\int\frac{d^2k}{(2\pi)^2}(\gamma_xk_x^2+\gamma_yk_y^4+\gamma_{nl}k^{\sigma-2}{k_y}^4)\hat{u}(\vec{k})\hat{u}(-\vec{k}),
\label{Heff}
\end{equation}
where $\mathcal{H}_0$ represents the energy of the unperturbed stripes configuration, $\hat{u}(\vec{k})$ is the Fourier transform of $u(\vec x)$ and the $\gamma$'s  are 
stiffness coefficients.  $\gamma_x$  measures the local elastic 
response of the stripes to compression or elongation of the pattern, while $\gamma_y$
is the local elastic response to bending of the stripes. $\gamma_{nl}$ comes from the 
long range repulsive interaction, which generates a non local interaction between far apart
pairs of layers, as illustrated in Fig.\ref{fig:stripes}.  Eq. (\ref{Heff}) in the $\sigma\to 2$ limit, is the usual elastic energy of 
a smectic phase\cite{deGPr1998,ChLu1995}.  
In Eq. (\ref{Heff}), it is assumed the existence of an ultraviolet cut-off $|\vec k| \ll k_0$, where the elastic theory makes sense.  
For $\sigma\geq2$, the non-local term results irrelevant in the long wave-length limit~\cite{MeStNi2015}. Therefore, in this case, we can fix $\sigma=2$. On the other hand, non-localities become 
important when $\sigma<2$. 
Thus, after rescaling of the stiffness coefficients, it is equivalent to work with the simpler effective Hamiltonian:
\begin{equation}
\Delta\mathcal{H}=
\frac{B}{2}\int\frac{d^2k}{(2\pi)^2}\left(k_x^2+l^2k^{\sigma-2}{k_y}^4\right)\hat{u}(\vec{k})\hat{u}(-\vec{k}),
\label{Heff1}
\end{equation}
which interpolates between the different possible cases. For $\sigma<2$, the long range tail of the repulsive 
interaction dominates at long wave-lengths and consequently the sigma depending term must be kept. On the other hand,
for fast enough decaying repulsive interaction ($\sigma\geq2$) the appropriate $k_y^4$ contribution is recovered by just fixing $\sigma=2$.       

Building upon the classical effective Hamiltonian (\ref{Heff1}), which has already been studied in \onlinecite{MeStNi2015}, our goal is to build a theory of quantum stripe melting,
where quantum fluctuations are considered on an equal footing as thermal fluctuations, and long range interactions are appropriately taken
into account. The well known classical theory of stripe melting~\cite{ToNe1981} has to be recovered as the high temperature limit
of the present formalism when $\sigma \geq 2$.

To promote the classical Hamiltonian (\ref{Heff1}) to the quantum realm it is necessary to introduce density fluctuations $\delta \phi$ in the form: 
\begin{equation}
\Delta \hat H=
\int\frac{d^2k}{(2\pi)^2} \left\{ \frac{1}{2\rho} |\delta \phi|^2+ \frac{B}{2}\left(k_x^2+l^2k^{\sigma-2}{k_y}^4\right)
|\hat{u}|^2 \right\}\;  ,
\label{Heff1-Quantum}
\end{equation}
where the parameter $\rho$ is the system {\em compressibility}\cite{Radzihovsky-2012}. The density fluctuations $\delta \phi$ and the displacement $\hat u$ are now operators acting on a Hilbert space.  In the absence of a time reversal or  a parity breaking field, such us a magnetic field, they can be considered as  canonically conjugate variables, $[\delta\phi(x), \hat u(x')]=i \delta(x-x')$. Equation 
  (\ref{Heff1-Quantum}) with canonical commutation relations completely defines a quantum model of stripe fluctuations. 
Although it is difficult to have an explicit expression for the compressibility $\rho$ in terms of more microscopic parameters, it is clear that it codifies quantum fluctuations of the system.  For instance, for  $\rho\to \infty$, the first term in equation (\ref{Heff1-Quantum}) vanishes. The result is a classical Hamiltonian. In the other incompressible  limit $\rho\to 0$, the energy contribution of density fluctuations is much larger that the displacement field, thus the system is in a deep quantum regime. In this way, the compressibility interpolates between a classical ($\rho\to \infty$) and a quantum regime ($\rho\to 0$).

To study a quantum system at finite temperature,  it is convenient to rewrite the theory in the imaginary time  coherent-state path integral formalism\cite{Sa2011}. After integration over $\delta \phi$, the action in terms of the displacement $\hat u$ reads, 
\begin{eqnarray}
\nonumber
\mathcal{S}&=&
\frac{B}{2}\int\limits_{\hat{u}(0)=\hat{u}(\beta)}d\tau \int\frac{d^2k}{(2\pi)^2}\left[\rho \, \partial_\tau\hat{u}(\vec{k},\tau)\partial_\tau\hat{u}(-\vec{k},\tau)\right.\\
&+&\left.(k_x^2+l^2k^{\sigma-2}{k_y}^4)\hat{u}(\vec{k},\tau)\hat{u}(-\vec{k},\tau)\right],
\label{qs1}
\end{eqnarray}
where we have conveniently rescaled the fields. The inverse temperature $\beta$ enters in the periodic boundary conditions of the displacement field along the imaginary time axes: $ \hat u(0,\vec k)=\hat u(\beta,\vec k)$. In this formalism, the density fluctuations are $\delta\phi\sim \rho\, \partial_\tau \hat u$.

Defining the Fourier transform of the field $\hat{u}(\vec{k},\tau)$ in the imaginary time direction in the form 
\begin{equation}
 \hat{u}(\vec{k},i\omega_n)=\frac{1}{\beta}\int_0^\beta d\tau e^{-i\omega_n\tau}\hat{u}(\vec{k},\tau),
\end{equation}
with the frequencies given by $\omega_n=2\pi n/\beta$ satisfying the periodicity condition in the interval $[0,\beta]$, the action (\ref{qs1}) 
can be recast as:  
\begin{eqnarray}
\nonumber
\mathcal{S}&=&
\frac{B\beta}{2}\int \frac{d^2k}{(2\pi)^2}\sum_n\left(\rho \omega_n^2+k_x^2\right.\\
&+&\left.l^2k^{\sigma-2}{k_y}^4\right)\hat{u}(\vec{k},i\omega_n)\hat{u}(-\vec{k},-i\omega_n).
\label{eq:qs2}
\end{eqnarray}
This is a Gaussian action for which the correlator is:
\begin{equation}
 \langle\hat{u}(\vec{k},i\omega_n)\hat{u}(-\vec{k},-i\omega_n)\rangle=\frac{(B\beta)^{-1}}{\rho \omega_n^2+k_x^2+l^2k^{\sigma-2}{k_y}^4}.
\end{equation}
From this we can obtain the mean square fluctuations (MSF) of translation and orientation degrees of freedom at a Gaussian level, 
which inform us on the stability of the possible phases in a system of stripes, depending on the range of 
the repulsive interactions $\sigma$, the strength of quantum fluctuations $\rho$ and temperature $\beta^{-1}$. The MSF of the displacement 
field $u$ are given by:
\begin{equation}
\langle u^2\rangle=\sum_{n=-\infty}^{\infty}\int\frac{d^2k}{(2\pi)^2}\frac{(B\beta)^{-1}}{\rho \omega_n^2+k_x^2+l^2k^{\sigma-2}{k_y}^4}.
\label{eq:u2}
\end{equation}
Additionally, considering that in the small
deviation limit the angular orientation of the stripe pattern is given by $\theta(\vec{r})=\partial_yu(\vec{r}))$,  we get for the MSF of the angular orientation: 
\begin{equation}
\langle \theta^2\rangle=\sum_{n=-\infty}^{\infty}\int\frac{d^2k}{(2\pi)^2}\frac{(B\beta)^{-1}k_y^2}{\rho \omega_n^2+k_x^2+l^2k^{\sigma-2}{k_y}^4}.
\end{equation}
For finite temperatures $\beta^{-1}>0$ one can conclude that positional order is always short ranged. 
Even at zero quantum fluctuations ($\rho\rightarrow\infty$), the 
zero frequency mode causes the MSF of the displacement field to diverge with the linear size of the system for any $\sigma$. On the other hand,
orientational order is always stable within this approximation for weak enough quantum fluctuations ($\rho\rightarrow\infty$) and low 
enough temperatures ($\beta^{-1}\rightarrow0$). Conversely, for high enough temperature or strong enough quantum fluctuations the orientational 
MSF grow monotonically without limit. Under such conditions some orientational phase transition is expected at intermediate values of $T$ and $\rho$. This picture 
may change (and indeed changes) under the effects of topological excitations~\cite{MeStNi2015}.  
The previous qualitative picture changes at zero temperature, since in this case it is possible to have long-range positional order. 
Indeed,  the  MSF of the displacement field, Eq. (\ref{eq:u2}),  becomes finite
for weak enough quantum fluctuations, independently of the value of $\sigma$. This is because in the limit $T \to 0$ the sum over Matsubara 
frequencies turns into an integral, which adds one effective dimension to the system, regularizing the integrals. On the other hand, for strong enough quantum fluctuations 
($\rho\rightarrow0$) the MSF of the displacement field grow without limit, observation which suggests the existence of a quantum phase transition at some
intermediate value of the strength of the quantum fluctuations. This is a transition from a positionally ordered smectic phase to a positionally disordered nematic phase.
The nature of the quantum smectic-nematic phase transition will be discussed in section \ref{Sec:Smectic}.
The qualitative picture just described is based on the elastic  action  (\ref{eq:qs2}) where only the contribution of small and smooth displacement fields were considered. 
The situation  changes if we also take into account the contribution of topological defects. In the following, we will
complete the description by the inclusion of  dislocations,
which  are essential to obtain qualitatively correct phase diagrams.

\begin{figure}[ht!]
\includegraphics[scale=0.5]{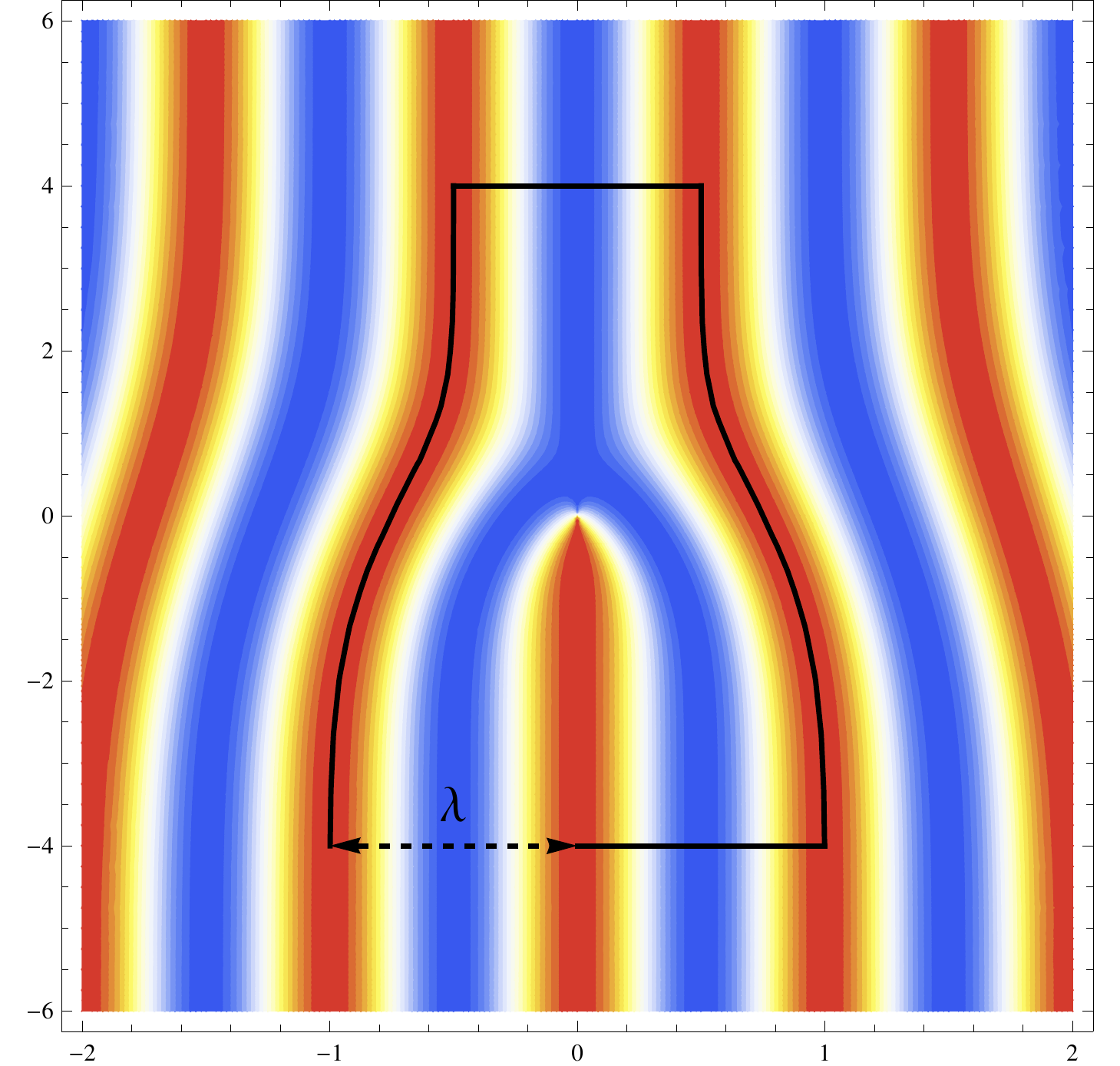}
\caption{Schematic representation of a dislocation. The drawn contour illustrates the jump in the value of the contour integral
$\oint\vec{\nabla}u\cdot d\vec{s}$ when the integration path encloses the dislocation core. 
The length of the dashed line corresponds to the value of the line integral. The parameter ``$\lambda$'' corresponds to the modulation length 
of the stripe pattern $2\pi/k_0$.}
\label{fig:dislocation}
\end{figure}

A dislocation in the kind of systems considered can be seen as a pair of stripes of
opposite densities coming to an end in the middle of the stripe pattern, as shown in Fig.\ref{fig:dislocation}.
  The strength of a dislocation is characterized by the line integral over a closed counterclockwise path around the dislocation core, as 
shown in Fig. \ref{fig:dislocation}. The double arrowhead signals the difference in the number of (pairs of) stripes at the top and the 
bottom of the drawn contour. 
This condition can be expressed in the form:
\begin{equation}
 \oint\vec{\nabla}u(\vec{r},\tau)\cdot d\vec{s}=z\lambda, \hspace{1cm} \forall \ \tau,
 \label{eq:cond1}
\end{equation}
where $z$ is an integer number and $\lambda$ stands for the modulation length (see Fig.\ref{fig:dislocation}) .
Here, it is important to note that we are dealing with a global neutral system, {\em i. e. }, $\int d^2 x\, \phi(x)=0$. 
For instance, if $\phi(x)$ represents a local magnetization perpendicular to the plane, we are dealing with {\em zero  global magnetization} phases. This implies that dislocations should be formed by an even number of stripes in order to keep neutrality. In this context, the parameter $\lambda$ represents the whole modulation length.
From a more technical point of view, we are dealing with phases that preserve global  $Z_2$ symmetry. The smectic and nematic phase transitions break translation and rotational invariance, however, they do not break the internal symmetry 
$\phi\to -\phi$.
Dislocations formed by an odd number of stripes violate this constraint. This fact has important consequences on the dynamic of the topological defects. For instance, odd stripe dislocations, if  they exist,  are forbidden to move along the stripe directions, due essentially to charge conservation, this effect is known as   ``glide constraint'' \cite{Nussinov2006}. In our case, we have no such a constraint since  even dislocations automatically
 preserve neutrality.

As discussed above, the action of Eq. (\ref{eq:qs2})  describes smooth fluctuations of the displacement field $u(\vec r,\tau )$.  In order  to write an effective action including the presence of dislocations, we split  the displacement field in the form 
$u(\vec r,\tau)=u_{\rm reg}(\vec r,\tau)+u_D(\vec r,\tau)$, where $u_{\rm reg}$ stands for the regular or smooth contribution and $u_{D}$  is the dislocation contribution.  Then, the total action can be written as 
\begin{equation}
\mathcal{S}[u]=\mathcal{S}_{\rm reg}[u_{\rm reg}]+\mathcal{S}_D[u_D] \;.
\end{equation}
$\mathcal{S}_{\rm reg}$ coincides with Eq. (\ref{eq:qs2}) and contains the smooth long wave-length deformations of the displacement $u(\vec r,\tau)$.  $\mathcal{S}_D $ is the dislocation contribution to the action that we will compute in the following.
The first step is to have an expression for a single dislocation profile.  At long distances, far away from the dislocation core,  the displacement $u(\vec r, \tau)$ should be {\em locally smooth}. Then, in order to compute the profile  we need to   minimize (\ref{eq:qs2}),  with  $u(\vec r,\tau)$  satisfying  the global  constraint (\ref{eq:cond1}). The Euler-Lagrange equation in momentum and frequency space amounts to: 
\begin{eqnarray}
\label{eq:Eu1}
\nonumber
&&-\frac{\delta\mathcal{S}}{\delta \hat{u}(-\vec{k},-i\omega_n)}=0,\\
&&-(\rho \omega_n^2+k_x^2+l^2{k_y}^4k^{\sigma-2})\hat{u}(\vec{k},i\omega_n)=0.
\end{eqnarray}
On the other hand, Eq. (\ref{eq:cond1}) implies that (through Stokes theorem) $\vec\nabla\times\vec\nabla u(\vec r,\tau)=z\lambda \delta^2(\vec r-\vec r')$, where $\vec r'$ is the position of the dislocation center. Thus, there is no smooth and single valued solution for this problem. Then, without loss of generality, we look for solutions with a discontinuity going from the dislocation center to infinity, for instance,  along the {\em y} axis (Fig.\ref{fig:dislocation}). In this way, the displacement field should satisfy $\lim_{x\to 0^+} u(x,y)-\lim_{x\to 0^-} u(x,y)=z\lambda$, for $y>0$ and  $0<\tau<\beta$.
The usual way to solve this problem\cite{ToNe1981,ChLu1995} is to introduce a singular source in the Euler-Lagrange equation that has the form $z\lambda \partial_x\delta(x-x_0(\tau))\Theta(y-y_0(\tau))$
in configuration space.  $x_0(\tau), y_0(\tau)$ is the position of the dislocation core, that can be generally considered a function of $\tau$.  Notice that this is a very localized source, {\em i.\  e.\  }, it is zero away from the  dislocation core and automatically implements the required discontinuity of the solution. Now, it is  immediate to solve Eq. (\ref{eq:qs2}) with the dislocation source in Fourier space. Turning back to configuration space we find,   
\begin{eqnarray}
\nonumber
u_D(\vec{r},\tau)&=&-\int_0^\beta\frac{d\tau'}{\beta}\int d^2\vec{r}' z\delta(\vec{r}'-\vec{r}_0(\tau'))\\ 
\nonumber
&\times&\int\frac{d^2\vec{k}}{(2\pi)^2}\sum_n e^{i\vec{k}\cdot(\vec{r}-\vec{r}')+i\omega_n(\tau-\tau')}\\
&\times&\lambda\frac{k_x}{k_y}\frac{1}{(\rho\omega_n^2+k_x^2+l^2{k_y}^4k^{\sigma-2})}\; ,
\end{eqnarray}
from which one can read the Green function for a single dislocation:
\begin{equation}
 G(\vec{k},i\omega_n)=-\lambda \frac{k_x}{k_y}\frac{1}{(\rho\omega_n^2+k_x^2+l^2{k_y}^4k^{\sigma-2})}.
 \label{eq:Gk1}
\end{equation}
Equation (\ref{eq:Gk1}), in the limit of $\omega_n=0$ and $\sigma=2$, coincides with the known result of a dislocation's Green function in classical  smectic phases\cite{ChLu1995,deGPr1998}.

The generalization of this result to the case of $N$ dislocations is straightforward. The topological constraint now reads:
\begin{equation}
\vec\nabla\times\vec\nabla u_D(\vec r,\tau)\vert_z=m(\vec r,\tau)=\lambda \sum_{n=1}^{N} z_n  \delta^2(\vec r-\vec r_n'(\tau))
\end{equation} 
where $z_n$ with $n=1,\ldots,N$ are arbitrary integers, and $\vec r'(\tau)$ are the trajectories of each of the $N$
dislocations cores. Moreover, we could also consider $m(\vec r,\tau)$ as a smooth function for a  {\em finite density of dislocations}. For an arbitrary distribution of dislocations, 
the displacement profile should be computed as:
\begin{equation}
u_D(\vec{r},\tau)=\int d\vec r' d\tau' \;  G(\vec r-\vec r', \tau-\tau') m(\vec r',\tau'),
\label{eq:uD}
\end{equation}
where $ G(\vec r-\vec r', \tau-\tau')$ is the Fourier transform of Eq. (\ref{eq:Gk1}). 
Taking into account  Eq. (\ref{eq:uD}) and Eq. (\ref{qs1}) we followed standard  procedures\cite{ToNe1981,Lubensky2000, ChLu1995,deGPr1998} to obtain $S_{D}[u_D]$ in terms of $m(\vec r,\tau)$. In Fourier space  we find:
\begin{eqnarray}
\nonumber
\mathcal{S}_D&=&
\frac{B\beta}{2}\int \frac{d^2k}{(2\pi)^2}\sum_n\left(\frac{1}{(\rho\omega_n^2+k_x^2)}\right.\\ \nonumber
&\times&\left.\frac{l^2\lambda^2k_x^2k_y^2k^{\sigma-2}}
{(\rho\omega_n^2+k_x^2+l^2{k_y}^4k^{\sigma-2})}+2E_da^2\right) \nonumber \\
&\times&\hat{m}(\vec{k},i\omega_n)\hat{m}(-\vec{k},-i\omega_n).
\label{eq:SD}
\end{eqnarray}
In this equation we have added the energetic contribution of a screened isolated dislocation, $E_d$, which is finite and it is not contained in the long wave-length  treatment.  
$a$ is short distance cut-off representing the dislocation core diameter. 

The correlation function of the angular variable $\theta(\vec{r})=\partial_y(u(\vec{r}))=\partial_y(u_{\rm reg}(\vec{r})+u_{D}(\vec{r}))$ is written  in Fourier space as:
\begin{equation}
 \langle\hat{\theta}(\vec{k},i\omega_n)\hat{\theta}(-\vec{k},-i\omega_n)\rangle=k_y^2\langle\hat{u}(\vec{k},i\omega_n)\hat{u}(-\vec{k},-i\omega_n)\rangle,
\label{eq:theta-u}
\end{equation}
and can be  computed from the actions Eq. (\ref{eq:qs2}) and  Eq. (\ref{eq:SD}) using Eq. (\ref{eq:uD}) to relate  $u_D(\vec r,\tau)$ with $m(\vec r,\tau)$.
The  leading order in the long wave-length ($k\rightarrow0$) and low temperature ($\omega_n\rightarrow0$) limits is:
\begin{eqnarray}
\nonumber
&&\langle\hat{\theta}(\vec{k},i\omega_n)\hat{\theta}(-\vec{k},-i\omega_n)\rangle=\\ 
&&\frac{k_BT}
{4E_da^2\rho\omega_n^2/\lambda^2+Bl^2k_y^2k^{\sigma-2}+2E_da^2k_x^2/\lambda^2}.
\label{eq:theta2}
\end{eqnarray}
This is a generalization of the corresponding results for classical smectic systems~\cite{ToNe1981}, taking into account the effects of long range repulsive interactions and quantum fluctuations.
From here, it is straightforward to infer the form of the corresponding orientational action:
\begin{eqnarray}
\nonumber
\mathcal{S}&=&
\frac{\beta}{2}\int \frac{d^2k}{(2\pi)^2}\sum_n\left(\frac{4E_da^2}{\lambda^2}\rho\omega_n^2+\frac{2E_da^2}{\lambda^2}k_x^2\right.\\
&+&\left.Bl^2k^{\sigma-2}{k_y}^2\right)\hat{\theta}(\vec{k},i\omega_n)\hat{\theta}(-\vec{k},-i\omega_n)\; .
\label{eq:so}
\end{eqnarray}
This action is not  rotationally invariant, since it represents angular fluctuations around a stripe pattern oriented along the $x$ axes. Thus, it is a good starting point  to study the stability of the orientational ordered phase under long wavelength  fluctuations. However,
in order to study critical properties of the stripe system,
we need to build the most general effective action that preserves local rotational symmetry, having the action (\ref{eq:so})
as the one obtained in the spin wave approximation. To this end, instead 
of tracking back the effects of the higher order contributions in the original action, which could be rather intricate, one can rebuild 
the full orientational Hamiltonian considering the symmetries of the stripe pattern. 
To identify the real symmetries of the system we should notice that the basic cell of any stripe system is composed by a double layer
of opposite densities with a small longitudinal width, as shown in Fig. \ref{fig:stripes}. In principle, any low energy configuration 
of the system of stripes can be built from these basic cells. In reference \onlinecite{MeStNi2015} a model which takes into account
the interactions between near and distant striped cells was introduced. In the next section we will discuss the qualitative aspects of
the model in more detail,
stressing the particular choice of observables for the identification of the different phase transitions.

\subsection{A plane rotors model}
\label{Sec:rotors}
The orientation of each basic cell can be defined by means of a local director vector $\vec N(\vec x) \equiv \vec\nabla\phi(\vec x)$. 
In general, the director $\vec N$ points perpendicular to the iso-density curves $\phi(\vec x)=\mbox{constant}$, and in the case of sharp interfaces it is peaked at the domain walls as shown in Fig. \ref{fig:stripes}. 
However, it is simple to realize that stripe configurations have no vectorial order since, integrating over the whole sample gives $\langle \vec N\rangle=0$. Notice that in two adjacent domain walls the directors always point in opposite directions. Moreover, if 
$\vec N$ is globally rotated by $\pi$, the final state is exactly de same as the original one.  Therefore, the orientational order is generally characterized by a quadratic function of $\vec N$, say a symmetric traceless tensor $Q_{ij}=N_iN_j-N^2\delta_{ij}/2 $, where the index $i,j =1,2$ refers to two orthogonal directions.  This is a nematic order parameter, which by construction is invariant under global rotations by $\pi$.
An orientationally ordered phase is characterized by $\langle Q\rangle\neq 0$, while for a completely disordered (isotropic) phase, $\langle Q\rangle= 0$. To completely characterize possible phase transitions it is necessary to compute fluctuations: $\langle Q_{ij}(x)Q_{\ell m}(x')\rangle$. From a technical point of view, this is a difficult calculation since it is a four point correlation function, $\langle Q_{ij}(x)Q_{\ell m}(x')\rangle\sim \langle \nabla_i\phi(x)\nabla_j\phi(x)\nabla_\ell\phi(x)\nabla_m\phi(x')\rangle$. From a physical point of view, this means that the nematic order parameter is not obtained as  a linear response to an homogeneous external field, but instead it is necessary to compute a quadratic response.  Within this approach, the minimum approximation to compute the nematic order parameter is the ``self consistent screening approximation'', where infinite sets of two-loop diagrams can be computed self consistently~\cite{BaMeSt2013}.   
\begin{figure}[ht!]
\includegraphics[scale=0.5]{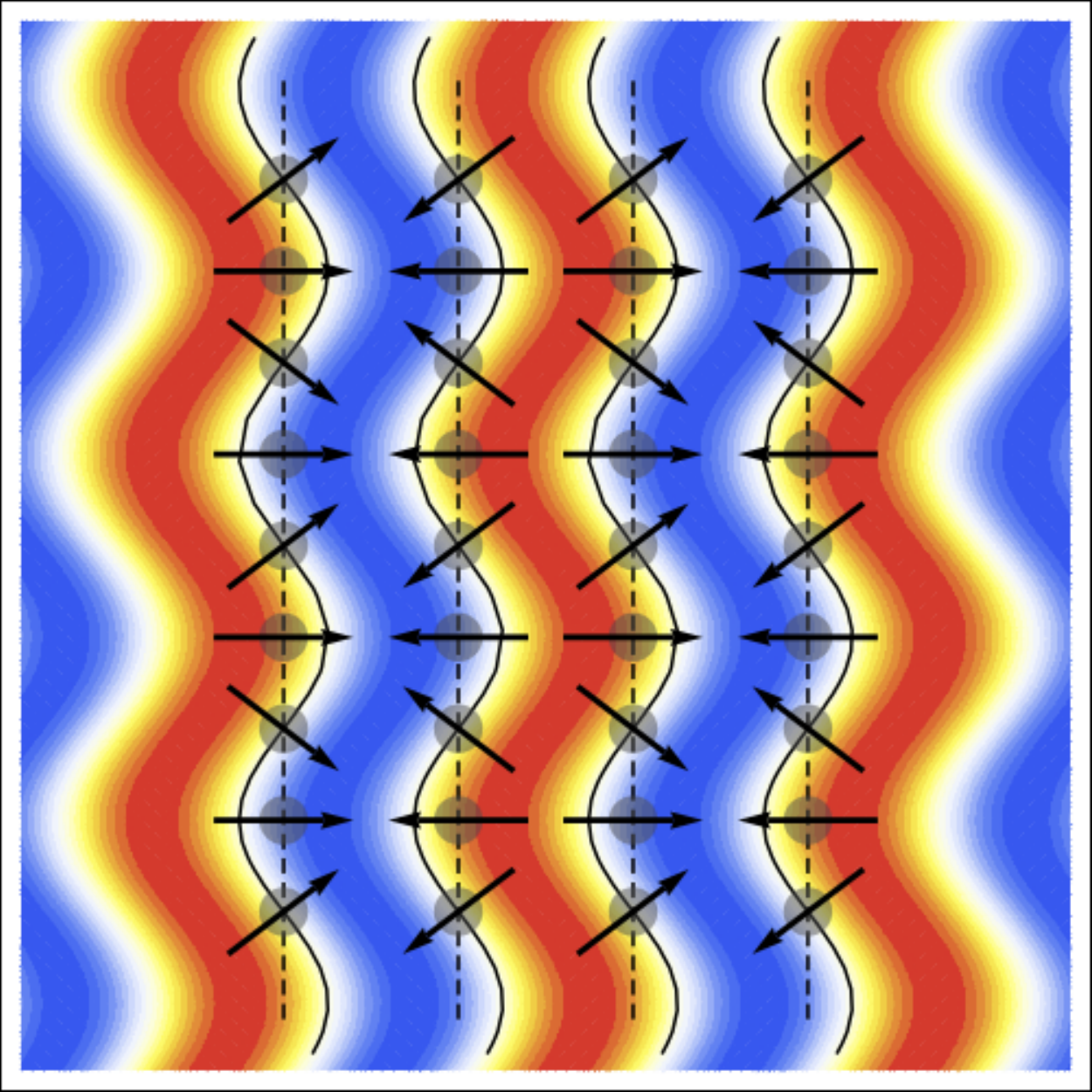}
\caption{The ``two sub-lattice'' structure of a stripe pattern. The black arrows indicate the local directors. Orientational order is
quantified by the mean orientation of the directors in one of the two sub-lattices. The small circles in the background illustrate the
reference state of the stripes were maximum orientational order is attained.}
\label{fig:sublattice}
\end{figure}
Here we propose an alternative observable to characterize the isotropic-nematic phase transition~\cite{MeStNi2015}.
Note that in the stripe phase, the directors are ordered in an ``antiferromagnetic'' structure.
Each block of two adjacent domain walls are characterized by two directors pointing in opposite directions and separated by half a stripe period.  Thus, the system is structured in two interpenetrated sub-lattices, both of them ferromagnetically ordered  with opposite directions. We can define a {\em sub-lattice director}  $\vec N_i$, where $i=a,b$ is the sub-lattice index. This is illustrated in Fig. \ref{fig:sublattice}. Then, similarly to what is done to describe antiferromagnetic order, it is useful to work with  two linear combinations of the sub-lattice directors 
\begin{eqnarray}
\vec N(x)&=&\frac{1}{2}\left(\vec N_a(x)+\vec N_b(x+\lambda/2)\right), 
\label{eq:director}  \\
\vec N_s(x)&=&\frac{1}{2}\left(\vec N_a(x)-\vec N_b(x+\lambda/2)\right),
\label{eq:stagger}
\end{eqnarray}     
where $\lambda=2\pi/k_0$ is the stripe period. 
In the stripe phase $\langle \vec N\rangle=(\langle \vec N_a\rangle+\langle \vec N_b\rangle)/2=0$, reflecting the fact that there is no dipolar or vector order. 
In turn, the staggered order parameter, Eq. (\ref{eq:stagger}), reflects the vectorial order of one of the sub-lattices,   $\langle \vec N_s\rangle=\langle \vec N_a\rangle$.   
It can be noted that if  $\langle N_s\rangle\neq 0$ the system is orientationally ordered. 
For the sake of  characterizing the isotropic-nematic phase transition one can work either with the tensor order parameter $Q_{ij}$ or with the vector one,  $\vec N_s$. Both of them characterize the same phase transition. On the other hand, the critical exponents can be quite different because $Q_{ij}$ and $\vec N_s$
are different observables. While $Q_{ij}$ is a quadratic response to an external uniform field, $\vec N_s$ is a linear response to a staggered conjugate field. In this sense, the latter is easier to compute since it is a linear function of the density gradient  $\vec\nabla\phi(x)$.  From the point of view of the symmetry, the nematic order is characterized by its invariance under rotations by $\pi$. This is obvious in the tensor order parameter since $\hat Q$ is a quadratic function of $\vec N$. On the other hand, $\vec N_s$ changes  sign under rotation by $\pi$. However, from Eq. (\ref{eq:stagger}) and Fig. \ref{fig:stripes} it is clear that a change of sign in $\vec N_s$ represents a change from high density (shaded areas) to low density (white areas) regions, and vice-versa. Then, $\vec N_s$ and $-\vec N_s$ represent the same state in the thermodynamic limit, as it should be.   
To describe the nematic transition using the staggered director we will assume that one sub-lattice is slaved to the other. Thus, we are not considering compression fluctuations, which can be treated perturbatively.

In the following, orientational order will be characterized by a space-time dependent unit director, given by 
\begin{equation}
\vec n(\vec x,\tau)=\frac{\vec N_a(\vec x,\tau)}{|\vec N_a|} \; .
\label{eq:n}
\end{equation}
The Hamiltonian for this order parameter can be expressed as the sum of two contributions: 
\begin{equation}
H[\vec n]=H_{sr}[\vec n]+H_{\ell r}[\vec n],
\label{eq:H}
\end{equation}
where $H_{sr}$ is a local function describing the effect of short-ranged interactions between the unit cells
and $H_{\ell r}$ codify the long-ranged interactions and it is in general non-local.
From symmetry considerations, the effective local Hamiltonian should have the form~\cite{ToNe1981,Lubensky2000,deGPr1998}:
\begin{equation}
H_{sr}=\frac{1}{2}\int d^2x \left[g_1(\vec{\nabla}\cdot\vec{n})^2 + g_2(\vec{\nabla}\times\vec{n})^2\right]\;, 
\end{equation}
since this is the more general quadratic form that is local and rotational invariant. Of course, it could contain non-quadratic local terms that we will treat perturbatively in a renormalization group approach. The coupling constants $g_1$ and $g_2$ will be defined later in terms of the stripe melting  parameters such as the stripe stiffness and dislocation's energy. 

To evaluate the contribution of the long ranged interactions we start by coarse graining the system, covering the  plane with small rectangles of area $A_i\sim \lambda a$ with  $a \ll\lambda$, centered at positions $x_i$ corresponding to a domain-wall sub-lattice as indicated in Fig. \ref{fig:stripes}.    
Considering two well separated rectangles of areas $A_1$ and $A_2$, located at a  distance $|x-x'| \gg \lambda$,
the interaction between them can be expressed in the form: 
\begin{equation}
\Delta H_{1,2}=\frac{1}{2}\int_{A_1} dx \int_{A_2}dx' \phi(x) J(x-x') \phi(x').
\end{equation}
Performing a multipolar expansion of the interaction $J(x)=J/|x|^\sigma$, summing over all possible  pairs of rectangles and retaining only the leading dipole contributions, the long range part
of the Hamiltonian can be written in the form:
\begin{eqnarray}
H_{\rm lr}&=&\frac{g}{2}\int d^2x\int d^2x' \Omega(\vert\vec{x}-\vec{x}'\vert)\left(\frac{\vec{n}(\vec{x})\cdot\vec{n}(\vec{x}')}
{\vert\vec{x}-\vec{x}'\vert^{\sigma+2}}\right. \nonumber \\
&-&(\sigma+2) \left.\frac{\vec{n}(\vec{x})\cdot(\vec{x}-\vec{x}')\vec{n}(\vec{x}')\cdot(\vec{x}-\vec{x}')}{\vert\vec{x}-\vec{x}'\vert^{\sigma+4}}\right), 
\end{eqnarray} 
where $g =\sigma J P^2$, the dipolar moment is given by  $P = \frac{1}{\lambda}\int_\lambda dx x\phi(x)$ and  $\Omega(x)$ is a short distance cutoff~\cite{MeStNi2015}.

Since we are interested in the classical as well as in the quantum behavior of this system, it is convenient to work in the Euclidean effective action formalism, in which we write the partition function as a functional integral of the form:
\begin{equation}
Z[\vec h,\beta]=\int {\cal D} \hat n \  e^{-S_g[\vec n(\vec x,\tau)]+\int d^2 x d\tau \, \vec h(\vec x)\cdot \vec n(\vec x)}
\end{equation} 
where the unitary vector $\hat n(\vec x,\tau)$ is a function of position and Euclidean time, satisfying the periodic boundary conditions $\hat n(\vec x,0)=\hat n(\vec x,\beta)$. $\vec h$ is the conjugate field of the order parameter. 
For static configurations, $\vec n(\vec x,\tau)\equiv \vec n(\vec x)$, $S_g[\vec n]=\beta H[\vec n]$, where $H$ is given by  Eq. (\ref{eq:H}). As described above, $H$ is strongly constrained by symmetry.  Conversely,  the order parameter dynamics cannot be deduced by means of symmetry properties only. 
As discussed in the previous section,
for the present case we consider a conservative local dynamics that preserves time reversal. 
Then, the Euclidean action can be written as:
\begin{eqnarray}
\nonumber
\mathcal{S}_g&=&\frac{1}{2}\int_0^{\beta}d\tau\int d^2x\ \left[g_0(\partial_\tau\vec{n})^2+ g_1(\vec{\nabla}\cdot\vec{n})^2 \right.\\ \nonumber
&+&\left. g_2(\vec{\nabla}\times\vec{n})^2\right]+\frac{g}{2}\int_0^\beta d\tau\int d^2x\int d^2x' \\ \nonumber
&\times&\,\Omega(\vert\vec{x}-\vec{x}'\vert)\left(\frac{\vec{n}(\vec{x},\tau)\cdot\vec{n}(\vec{x}',\tau)}
{\vert\vec{x}-\vec{x}'\vert^{\sigma+2}}\right. -(\sigma+2)\\
 &\times& \left.\frac{\vec{n}(\vec{x},\tau)\cdot(\vec{x}-\vec{x}')\vec{n}(\vec{x}',\tau)\cdot(\vec{x}-\vec{x}')}{\vert\vec{x}-\vec{x}'\vert^{\sigma+4}}\right).
\label{eq:Sg}
\end{eqnarray}
$S_g[\vec n]$ describes the orientational order properties of a stripe forming system due to competing interactions at different scales. $\vec n$  represents the director of one sub-lattice, as described above, and $\langle \vec n\rangle $ is the linear response to a staggered conjugate field with a periodicity $\lambda$. In Eq. (\ref{eq:Sg}) there are three local terms, given by the coupling constants $g_0, g_1,g_2$. $g_0$ measures the intensity of the quantum fluctuations, while $g_1$ and $g_2$ are stiffness coefficients.  The effects of long-ranged interactions are contained in the last two non-local terms, both proportional to the coupling constant $g$. Notice that these terms are a generalization of a dipolar interaction, with an isotropic component as well as an  anisotropic one.
 Both components decay as $1/|x|^{\sigma+2}$. It can be shown that the Euclidean action (\ref{eq:Sg}) matches the one obtained in the spin wave limit, eq. (\ref{eq:so}), when the coupling constants are given by:
\begin{eqnarray}
\label{eq:g0}
g_0&=&\frac{4E_da^2}{\lambda^2}\rho\\ \nonumber
g_1&=&\gamma_y\\ \nonumber
g_2&=& \frac{2E_da^2}{\lambda^2}\\ \nonumber
g&=&c\gamma_{nl}.
\end{eqnarray}
The classical isotropic-nematic transition is governed by the fixed point $g_1=g_2$. Any small deviation from this situation will flow towards the fixed point when the renormalization group is implemented\cite{ChLu1995}.
Since at hight temperatures we should  reproduce the classical result, we expect that, at
length scales much longer than the modulation length, the anisotropy in the spatial stiffnesses
vanishes. This leads to a simpler effective action given by
\begin{eqnarray}
\nonumber
\mathcal{S}_g&=&\frac{1}{2}\int_0^{\beta}d\tau\int d^2x\ \left[g_0(\partial_\tau\vec{n})^2+ g_1 \sum_{\mu=1,2}|\partial_\mu\vec{n}|^2\right] \\ \nonumber
&+&\frac{g}{2}\int_0^\beta d\tau\int d^2x\int d^2x'\,\Omega(\vert\vec{x}-\vec{x}'\vert)\left(\frac{\vec{n}(\vec{x},\tau)\cdot\vec{n}(\vec{x}',\tau)}
{\vert\vec{x}-\vec{x}'\vert^{\sigma+2}}\right. \\
 &-& \left.(\sigma+2)\frac{\vec{n}(\vec{x},\tau)\cdot(\vec{x}-\vec{x}')\vec{n}(\vec{x}',\tau)\cdot(\vec{x}-\vec{x}')}{\vert\vec{x}-\vec{x}'\vert^{\sigma+4}}\right).
\label{eq:Soq}
\end{eqnarray}

As anticipated in Section \ref{Sec:Melting}, at $T=0$ there can be two different kind of orders: positional  and orientational. However, only
orientational order survives upon inclusion of thermal fluctuations. Thus, in the following sections we will consider first the zero
temperature phase transitions and then the finite $T$ phase diagram. 
In Section \ref{Sec:Criticality} we will study the isotropic-nematic quantum phase transition. Then, in Section \ref{Sec:Smectic}
the smectic-nematic quantum phase transition will be analysed. Finally, in Section \ref{Sec:Thermal} we will consider the effects of thermal
fluctuations and complete the phase diagrams of the system.

\section{Nematic order at zero temperature:  Quantum Criticality}
\label{Sec:Criticality}
It has been shown~\cite{MaSc2004} that the universality class  does not change if the anisotropic dipolar 
interaction is replaced by an attractive isotropic term, decaying with the same power law as the original interaction. Then, 
as far as universality is concerned, the action can be further simplified to read:
\begin{eqnarray}
\label{eq:Solr}
\lefteqn{
\mathcal{S}_{\rm glr}=\frac{1}{2}\int_0^{\beta}d\tau\int d^2x\ \left[g_0(\partial_\tau\vec{n})^2+ g_1 \sum_{\mu=1,2}|\partial_\mu\vec{n}|^2\right]} \\ \nonumber 
&-&\frac{g}{2}\int_0^\beta d\tau\int d^2x\int d^2x'\,\Omega(\vert\vec{x}-\vec{x}'\vert)
\frac{\vec{n}(\vec{x},\tau)\cdot\vec{n}(\vec{x}',\tau)}
{\vert\vec{x}-\vec{x}'\vert^{\sigma+2}}\;, 
\end{eqnarray}
which is a  non-linear sigma model with long range interactions. 
As we have discussed in previous sections, we expect a phase transition controlled by $g_0\sim \rho$ (see eq. (\ref{eq:g0})). Thus, it should exist a critical value $g^c_0\sim \rho_c$ separating an ordered phase ($g_0>g_0^c$) from a quantum disordered one for $g_0<g_0^c$.
To compute critical properties, instead of working with a non-linear sigma model, it is simpler to work with the linear sigma model, since both models belong to the same universality class, provided the order parameters share the same symmetry properties and the interactions of both models have the same long wave-length behavior.\cite{Sa2011,Brzi1976,BrLe1976}
There is an intuitive way to understand this fact. It is possible to eliminate $g_0$ and $g_1$ from de action, by just rescaling space-time coordinates and the field $\vec n$. The result is an isotropic rotor model in three dimensions, with $\vec n\cdot\vec n= g_0^{1/2} g_1$.  We see that the net effect of $g_0$ is to control the modulus of the vector field $\vec n$. Thus, the phase transition can be tuned by controlling the strength of the vector field. Then, we can relax the constraint of fixed modulus by adding, instead, a potential, $V(|\vec n|^2)$, with deep minima at  $\vec n\cdot\vec n= g_0^{1/2} g_1$. 
We can write, 
\begin{eqnarray}
\nonumber
\lefteqn{
\mathcal{S}_{glin}=\frac{1}{2}\int_0^{\beta}d\tau\int d^2x\ \left[(\partial_\tau\vec{n})^2+ \sum_{\mu=1,2}|\partial_\mu\vec{n}|^2\right. } \\ \label{eq:Sglin}
&+&\left. r\,\vec{n}(\vec{x},\tau)^2+\frac{2u}{4!}\vec{n}(\vec{x},\tau)^4 \right] \\ \nonumber
&-&\frac{\tilde g}{2}\int_0^\beta d\tau\int d^2x\int d^2x'\,\Omega(\vert\vec{x}-\vec{x}'\vert)
\frac{\vec{n}(\vec{x},\tau)\cdot\vec{n}(\vec{x}',\tau)}{\vert\vec{x}-\vec{x}'\vert^{\sigma+2}},
\nonumber
\end{eqnarray}
where the restriction on the modulus of the vector $\vec n$ has been lifted. 
Eqs. (\ref{eq:Solr}) and (\ref{eq:Sglin}) represent very different models. The former is a rotor model, {\em i.\  e.\  } there is a  hard constraint on the modulus of the vector field. On the other hand, the latter is an $O(2)$ model without any constraint. The connection between both models reside in the fact that  the degrees of freedom of the non-linear sigma model (Eq. (\ref{eq:Solr})) are the Goldstone modes of the linear model (eq. (\ref{eq:Sglin})) in the broken symmetry phase where  $\langle\vec n\cdot\vec n\rangle= -6r/u=g_0^{1/2} g_1$. The longitudinal fluctuations are gapped, being irrelevant in the renormalization group sense. Thus, both models, even having quite different behaviors, share the same critical properties,  i.e. they are in the same universality class. The equivalence of  both models at criticality can be rigorously shown by  using a Hubbard-Stratonovich transformation to lift the constraint\cite{Amit1978}. Very near  the critical point, it can be shown that $(r-r_c)/r_c\sim (g_0^c-g_0)/g_0^c=(\rho_c-\rho)/\rho_c$, where $r_c$, $g_0^c$ and $\rho_c$ are the corresponding critical values and in the last equality we have used eq. (\ref{eq:g0}).  In this way, the parameter $r$ in the linear model, controls the mean value of the vector modulus, in the same way that $g_0$ does in the non-linear sigma model. Therefore, the parameter $r$ controls quantum fluctuations equivalently to the compressibility $\rho$. However,  increasing values of $r$ correspond to decreasing values of $\rho$. Thus, the classical limit is described by 
$r \ll r_c$, while the strong quantum regime corresponds to  $r \gg r_c$.

 To analise critical properties, 
we found convenient to re-write Eq. (\ref{eq:Sglin}) for an $N$-dimensional vector $\vec n(\vec x,\tau)$  in Fourier space for general spatial dimension $d$  and  a long ranged interaction decaying as $|\vec x-\vec x'|^{-(\sigma+d)}$:
\begin{eqnarray}
\nonumber
\mathcal{S}_{o}&=&\frac{1}{2}\sum_n\int \frac{d^dk}{(2\pi)^d}\ (\omega_n^2+k^{\sigma}+r) \,
\hat{n}_{\alpha}(\vec{k},\omega_n)\hat{n}_{\alpha}(-\vec{k},-\omega_n) \\  
&+&\frac{u}{4!}\int_0^{\beta}d\tau\int d^dx\ \vec{n}(\vec{x},\tau)^4,
\label{eq:SoGL1}
\end{eqnarray}
where  $\omega_n$ are the Matsubara frequencies  and a sum over the vector component indexes $\alpha=1\ldots N$  is understood. We recall that in Eq. (\ref{eq:SoGL1}), in a similar way than Eq. 
(\ref{Heff1}), if $\sigma\geq 2$, we should consider $k^2$ as the dominant contribution to the dispersion. On the other hand, for 
$\sigma< 2$, $k^\sigma$ is the dominant term.
By fixing $d=2, N=2$, we  recover the original model.  The critical properties of the model (\ref{eq:SoGL1}) at $T=0$ have been studied by means of the epsilon expansion technique in reference \onlinecite{DuBh2001}. In the following subsections we summarize the results, offering an alternative way to compute critical exponents, specially useful at finite temperature which we present in the next section. 

As we have already discussed, the behavior of positional and orientational fluctuations allow us to classify  the models according to the range of the interaction,
whether  $\sigma \geq 2$ or $\sigma <2$.

\subsection{$\sigma \geq 2$}
When $\sigma \geq 2$ the interactions are effectively short ranged and the non-local terms in  (\ref{eq:Sglin}) are irrelevant in the renormalization group sense.  
Then, one can simply fix $\sigma=2$ in Eq. (\ref{eq:SoGL1}), which at $T=0$ reads: 
\begin{eqnarray}
\nonumber
\mathcal{S}&=&\frac{1}{2}\int \frac{d\omega d^dk}{(2\pi)^d}\ (\omega^2+k^{2}+r) \,
\hat{n}_{\alpha}(\vec{k},\omega)\hat{n}_{\alpha}(-\vec{k},-\omega) \\  
&+&\frac{u}{4!}\int_0^{\infty}d\tau\int d^dx\ \vec{n}(\vec{x},\tau)^4\;.
\label{eq:SlocalT0}
\end{eqnarray}
This is the well known classical $O(N)$ model in effective dimension $d_{\rm eff}=d+1$~\cite{Amit1978}. Its critical behavior is very well established and very good approximations for the critical exponents are known.  
The upper critical dimension is $d_{\rm eff}=4$, or equivalently the spatial upper critical dimension is $d_u=3$. While for $d_u>3$ the Gaussian fix point is stable and  mean field gives the correct critical behavior, for $d=2$ it is necessary to make 
an $\epsilon=3-d$ expansion. It turns out that there is a non-trivial fix point of order $\epsilon$  at $r=r_2^c=-\epsilon(n+2)/(n+8)$  and $u=u_2^c=4 \epsilon/(n+8)$. 
The susceptibility, $\chi\sim |r-r_c|^{-\gamma}$, and correlation length, $\xi\sim (r-r_c)^{-\nu}$, are characterized by the exponents
\begin{eqnarray}
\gamma&=&1+\frac{1}{2}\left(\frac{N+2}{N+8}\right)\left(3-d\right) \label{eq:gamma-local} \\
\nu&=&\frac{1}{2}+\frac{1}{4}\left(\frac{N+2}{N+8}\right)\left(3-d\right). \label{eq:nu-local}
\end{eqnarray}

\begin{figure}[ht!]
\includegraphics[scale=0.4]{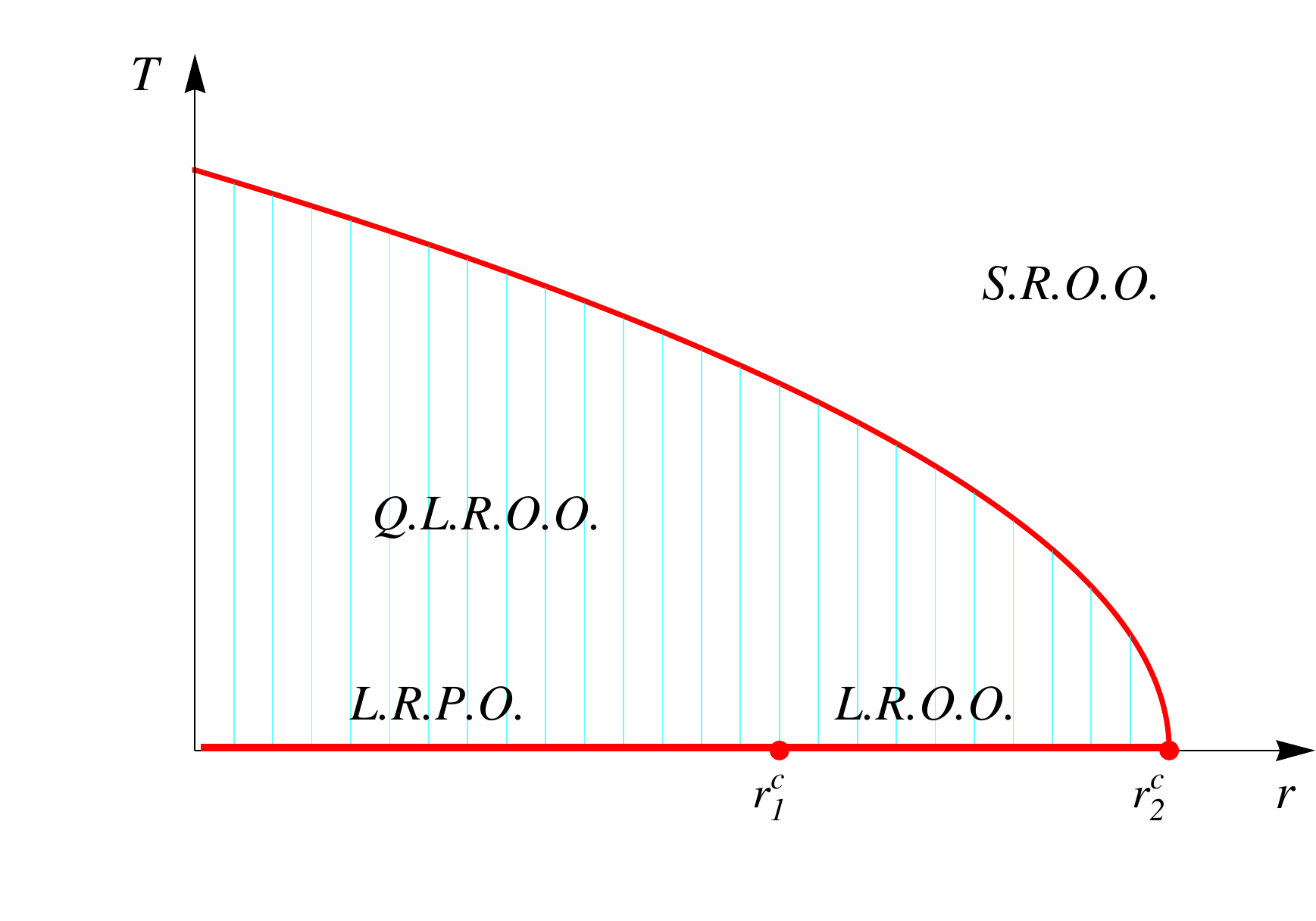}
\caption{(Color online) Qualitative phase diagram for the orientational order parameter for  $\sigma\geq2$ in $d=2$. 
$r_1^c$ marks a $T=0$ first order phase transition between a phase with long range positional order (L.R.P.O) and a long
range orientationally ordered one (L.R.O.O). A quantum critical
point at $r_2^c$ separates the orientational long range ordered phase from a quantum disordered one also called short ranged orientational ordered (S.R.O.O). 
At finite temperature only  quasi long range orientational order (Q.L.R.O.O) is present ending at a KT critical line. }
\label{fig:phasediag1}
\end{figure}

In the special case of interest in this work, $N=2$, $d=2$ at $T=0$ is equivalent to a classical  XY model in three dimensional space at finite temperature. The temperature in the classical system plays the role of the inverse coupling constant in the quantum system at $T=0$\cite{HoCa2014}.  It is known that the classical model has a critical point driven by spin wave fluctuations at a finite $T_c$ and several critical exponents for the $d=3$ XY universality class have been computed numerically with great precision~\cite{CaHa2001}. These results are compatible with more recent ones on the quantum $d=3$ $O(2)$ model~\cite{Langf2013}. This quantum critical point (QCP) is depicted in Figure \ref{fig:phasediag1}.

\subsection{$\sigma < 2$}
In this case the generalized dipolar interaction, last term of Eq. (\ref{eq:Sglin}),  is generally a relevant interaction depending on the space dimension $d$. 
Starting  from the action (\ref{eq:SoGL1}), the susceptibility at one loop order reads:
 \begin{equation}
 \label{eq:chi}
 \chi(k,\omega,d,\sigma,r,u)^{-1}=k^\sigma+\omega^2+r+2\frac{u}{4!}N\langle\hat{n}_{\alpha}^2\rangle,
\end{equation}
 where $\langle\hat{n}_{\alpha}^2\rangle$ is determined self-consistently from
\begin{equation}
\langle\hat{n}_{\alpha}^2\rangle=\int\frac{d^dk\ d\omega}{(2\pi)^{d+1}}\frac{1}{k^\sigma+\omega^2+r+2\frac{u}{4!}N\langle\hat{n}_{\alpha}^2\rangle}.
\label{eq:nalpha}
\end{equation}
By performing  the change of variables:
\begin{eqnarray}
 k'&=&k^{\sigma/2}, \nonumber\\
 d'&=&\frac{2}{\sigma}d, \nonumber\\ 
 u'&=&uf(d,\sigma),
 \label{eq:relations}
 \end{eqnarray}
 where $f(d,\sigma)=(2\pi)^{d-d'}(2/\sigma) (S_d/S_{d'})$ and  $S_d$  is the area of $d$-dimensional sphere,
 we find that  
\begin{equation}
 \chi(k,\omega,d,\sigma,r,u)=\chi(k',\omega,d',2,r,u'),
 \label{eq:chirelation}
\end{equation}
where $\chi(k',\omega,d',2,r,u')$ is the susceptibility of the short range interacting case ($\sigma=2$), computed from eq. (\ref{eq:SlocalT0}), with renormalized values of $k$, $u$ and $d$.
This relation allows us to compute the critical exponent  of the long-ranged interacting models from the knowledge of the local models, at least at the one loop approximation. For instance, 
the critical exponent of the susceptibility should satisfy  $\gamma(\sigma,d)=\gamma(2,d')$. 
Then, using eq. (\ref{eq:gamma-local})  we immediately find, 
\begin{equation}
 \gamma(\sigma,d)=1+\frac{1}{\sigma}\frac{(N+2)}{(N+8)}\left(\frac{3\sigma}{2}-d\right),
\end{equation}
which coincides with the result obtained by a direct calculation within the perturbative renormalization group at linear order in an $\epsilon$-expansion\cite{DuBh2001}.

The upper critical dimension $d_u$ for $\sigma<2$ is obtained from the well known value $d'_u=3$ for $\sigma=2$,  and the scaling relations of eq.
(\ref{eq:relations}), giving $d_u=3\sigma/2$. We note that the upper critical dimension depends continuously on $\sigma$. This is a direct consequence of the fact  that the dynamical exponent $z=\sigma/2$ also depends on $\sigma$, since the dispersion relation of the lowest energy modes is $\omega\sim k^z\sim k^{\sigma/2}$. Of course, $z\to 1$  when we take the limit  $\sigma\to 2$.  

A second critical exponent for $\sigma<2$  can be obtained by
noticing that the scaling of the wave vectors  $k=(k')^{2/\sigma}$ implies that any length scale
$\ell$ in the original system
is related to the length scale $\ell'$ in the local $\sigma=2$ system by $l=(l')^{2/\sigma}$. In particular,  the correlation length should satisfy $\xi=(\xi')^{2/\sigma}$, implying that  the associated critical exponent  will be given by $ \nu(\sigma,d)=\frac{2}{\sigma}\nu(2,d')$. Using eq. (\ref{eq:nu-local}) and the scaling relations of eq. (\ref{eq:relations}), we find
\begin{equation}
 \nu(\sigma,d)=\frac{1}{\sigma}\left[1+\frac{1}{\sigma}\frac{(N+2)}{(N+8)}\left(\frac{3\sigma}{2}-d\right)\right],
\end{equation}
which coincides with the expression computed by means of a linear expansion in $\epsilon=3\sigma/2-d$~~\cite{DuBh2001}.   
It is interesting to note that, the larger the interaction range, the smaller the upper critical dimension. As a consequence, for $\sigma<2d/3$, the exponents coincide with the mean-field ones $\gamma=1$, $\nu=1/\sigma$. In particular, these exponents are exact for the Coulomb interaction 
($\sigma=1$) in $d=2$.  

\begin{figure}[ht!]
\includegraphics[scale=0.4]{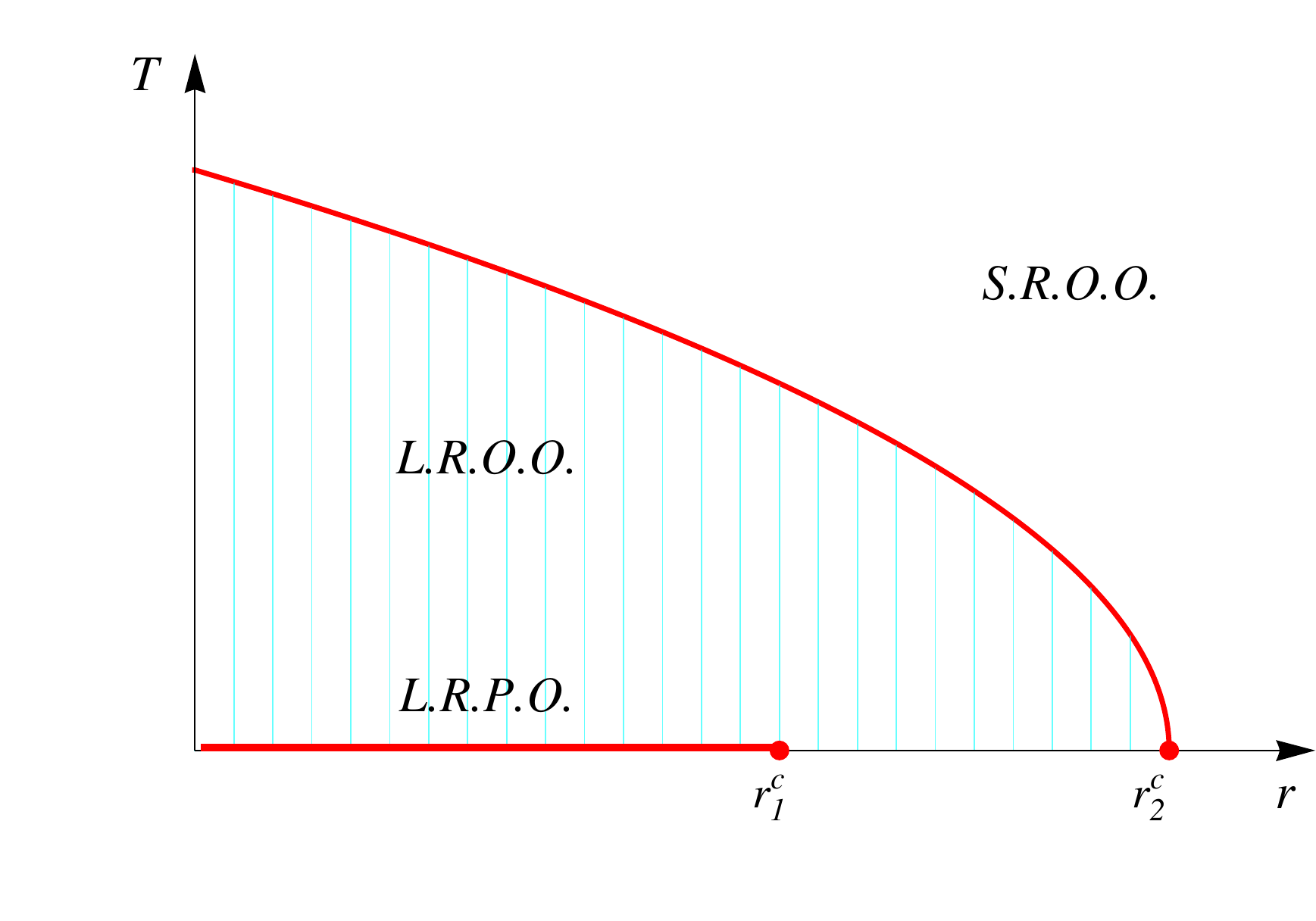}
\caption{(Color online) Qualitative phase diagram for the orientational order parameter for $\sigma<2$ in $d=2$. 
$r_1^c$ marks a $T=0$ phase transition between a phase with long range positional order (red line)  (L.R.P.O) and a long
range orientational ordered one (L.R.O.O). For $4/3 < \sigma \leq 2$ the transition is of first order induced by
quantum fluctuations, while for $\sigma \leq 4/3$ it is of second order. 
A quantum critical
point at $r_2^c$ separates the long range orientational  ordered phase (L.R.O.O) from a quantum disordered one or short ranged orientational ordered (S.R.O.O). 
The finite temperature critical line (red) separates 
a long range ordered phase from a disordered one, ending at the QCP. }
\label{fig:phasediag2}
\end{figure}

Once two critical exponents are known, other ones like $\eta,\alpha,\beta$ can be immediately obtained by scaling and hyperscaling relations\cite{Mucio2001}:
\begin{eqnarray}
\nu(2-\eta) &=&\gamma \label{eq:scaling1}\\
\alpha+2\beta+\gamma&=&2  \label{eq:scaling2}\\
2-\alpha&=&\nu (d+z), \label{eq:hyperscaling}
\end{eqnarray}
taking into account that, at one loop approximation, $z=\sigma/2$.
In table \ref{tb:exponents} we summarize the results obtained  for different  values of the interaction range parameter $\sigma$.
The corresponding phase diagram is shown in Figure \ref{fig:phasediag2}.
\begin{table*}
\begin{center}
\[
\begin{array}{||l|c|c|c|c||}
\hline\hline
 &   &  \multicolumn{2}{c|}{\mbox{Long-range interactions}}  & \mbox{Short-range interactions}   \\
 \hline 
 & \mbox{~~~}&  \sigma\le \frac{2d}{3}&\frac{2d}{3}< \sigma< 2  & \sigma\ge 2 \\
 \hline
 & \gamma &  1 &   1+\frac{1}{\sigma}\left(\frac{N+2}{N+8}\right)\left(\frac{3\sigma}{2}-d\right) & 
1+\frac{1}{2}\left(\frac{N+2}{N+8}\right)\left(3-d\right) \\
\cline{2-5}
T=0  & \nu&  \frac{1}{\sigma} &   \frac{1}{\sigma}+\frac{1}{\sigma^2}\left(\frac{N+2}{N+8}\right)\left(\frac{3\sigma}{2}-d\right) & 
\frac{1}{2}+\frac{1}{4}\left(\frac{N+2}{N+8}\right)\left(3-d\right) \\
\cline{2-5}
& \eta &  2-\sigma & 2-\sigma  &   0 \\
\cline{2-5}
& \beta &  \frac{1}{2} & \frac{1}{2}\left[1-\frac{1}{\sigma}\left(\frac{3\sigma}{2}-d\right)\right] & 
   \frac{1}{2}\left[1-\frac{1}{2}\left(3-d\right)\right] \\
\cline{2-5}
& \alpha & 0 & \frac{1}{\sigma}\left(\frac{3\sigma}{2}-d\right)
\left[ 1-\frac{1}{\sigma}\left(\frac{N+2}{N+8}\right) \left(d+\frac{\sigma}{2}\right)\right] &
\frac{\left(3-d\right)}{2}
\left[ 1-\left(\frac{N+2}{N+8}\right) \frac{\left(d+1\right)}{2}\right]   \\
\hline\hline 
 &   \xi(T, r=r_c) &  T^{-\frac{1}{\sigma}\left(\frac{2d}{\sigma}-1\right)} &  T^{-\frac{2}{\sigma}} & T ^{-1} \\
 \cline{2-5}  
T\neq 0 &   \chi(T, r=r_c) &  T^{-\left(\frac{2d}{\sigma}-1\right)} &  T^{-2} &  T^{-2} \\
 \cline{2-5} 
&   \delta r(T)=r_c-r & -T^{\left(\frac{2d}{\sigma}-1\right)}   & -T^{2}  &  -T^{2} \\
 \hline\hline 
\end{array}
\]
\caption{Critical properties of the $O(N)$ model with generalized dipolar interactions, (Eq. (\ref{eq:SoGL1})), at fixed $d\lesssim3$.
For $T=0$ the critical exponents $\gamma,\nu,\eta,\alpha,\beta$ are shown as a function of the interaction range parameter $\sigma$. There are two main columns: models with short-ranged interactions ($\sigma\ge 2$) and long ranged interactions ($\sigma< 2$). The short ranged case (last column) shows the exponents of the local $O(N)$ model in $d+1$ dimensions,  computed at linear order in an $\epsilon$-expansion around the upper critical dimension $d=3$\cite{Amit1978}. In the long-ranged case $\sigma<2$, the upper critical dimension is $d_u=3\sigma/2$ leading to two different regimes. In the first column, $\sigma<2d/3$, the system is always above the upper critical dimension and then the exponents are dominated by the Gaussian fixed point  and are exact. In the second column, we show the critical exponents computed by means of the scaling properties of the susceptibility (eq. (\ref{eq:relations})) which coincide with a direct perturbative RG calculation\cite{DuBh2001}.  $\eta,\alpha,\beta$ have been computed invoking the scaling and hyperscaling relations of Eqs. (\ref{eq:scaling1}-\ref{eq:hyperscaling}).
The lower part of the table displays the finite T behavior of the correlation length and the susceptibility at the critical point for $T\to 0$.  The last column shows the  results for the local model\cite{Sa2011}. The first two columns display the results of the long-ranged interacting model computed from the scaling behavior of Eq.  (\ref{eq:relations}). The behavior of the critical line $\delta r(T)$ very near the QCP is also shown.
\label{tb:exponents}}
\end{center}
\end{table*}

\section{Smectic-Nematic quantum  phase transition}
\label{Sec:Smectic}
Although positional order is destroyed at any finite temperature already at the level of Gaussian fluctuations, as shown in section
\ref{Sec:Melting}, it can actually exist at $T=0$. 
From Eqs. (\ref{eq:theta-u}) and (\ref{eq:theta2}) it can be inferred that, at $T=0$, the proliferation of unbounded dislocations
destroys positional order in a region where orientational order is still possible. Then, besides the isotropic-nematic transition
described in section \ref{Sec:Criticality}, the system also displays a nematic-smectic quantum phase transition. 

A theory of the  thermal  nematic-smectic phase transition was developed by McMillan and de Gennes\cite{McMi1971,deGPr1998}. Some
generalizations to quantum systems were considered in Refs. \onlinecite{Lawler2008,Lawler-Erratum2009,BarciFradkin2011}. Here, we follow
similar procedures to understand the nature of the  smectic-nematic quantum phase transition as a function of the interaction 
range parameter $\sigma$.  
The theory starts from an orientationally ordered state characterized by the director $\vec n(x)$,  pointing for instance 
in the $x$ direction:  $\vec n(x)=n_x \^ i$. On top of this we consider a positional order parameter given by a complex function 
$\psi(\vec{x},\tau)=\rho(\vec{x},\tau)\exp(-i k_0 u(\vec{x},\tau))$.  $\rho$ is the smectic order parameter,  indicating the presence
of a density modulation. The phase is proportional to the displacement field  $u(\vec{x},\tau)$. To build a rotationally invariant 
action, the smectic order parameter and the nematic fluctuations should be coupled by a ``covariant derivative''
$D_x=\partial_x$ and   $D_y=\partial_y-ik_0n_y(\vec{x},\tau)$ ($x$ is the coordinate along the director $\vec n$, while the 
coordinate $y$ is transversal). Then, the action for the smectic order parameter coupled to nematic fluctuations is given by: 
\begin{eqnarray}
\mathcal{S}\left[\psi,n_y\right]&=&\int d^2x\ d\tau\left\{    
\frac{a}{2}\mid\psi\mid^2 +\frac{b}{4}\mid\psi\mid^4    \right. 
\label{eq:MdG} \\ 
&+&\left. \frac{C_x}{2}\left|\frac{\partial \psi}{\partial x}\right|^2+\frac{C_y}{2}\left|\left(\frac{\partial }{\partial y}-ik_0n_y\right)\psi\right|^2\right\} \nonumber\\ 
&+&\int\frac{d^2k}{(2\pi)^3}d\omega
\left(g_0\omega^2+ g k^\sigma\right)
\hat{n}_y(\vec{k},i\omega)\hat{n}_y(-\vec{k},-i\omega)\; .
\nonumber
\end{eqnarray}
In this equation, $a$ controls the phase transition, while $b>0$. $C_x$ and $C_y$ are  smectic elastic constants. Since the 
system is anisotropic,  they are generally  different.  The last term of Eq. 
(\ref{eq:MdG}) represents the nematic dynamics,  where $g_0$ and $g$ are couplings associated to the small angular fluctuations
of the director, which can be traced back 
to the action (\ref{eq:Solr}). Eq. 
(\ref{eq:MdG})  resembles  the effective free energy for smectic-nematic  transition in 3D liquid crystals and the metal-superconductor
transition in type 1 
superconductors~\cite{McMi1972,McMi1973,deGPr1998}. 

From the first line of Eq. (\ref{eq:MdG}), it is simple to conclude that, in the absence of nematic fluctuations, the phase
transition is second order, driven by the sign of  the parameter $a$. However, nematic fluctuations can change the nature 
of the transition depending on the range of the interactions. To see this, we can integrate out the Gaussian transverse nematic
fluctuations in $n_y$. This leads us to an effective action $\mathcal{S}_{ef}[\psi]$, such that:
\begin{equation}
\exp(-\mathcal{S}_{ef}[\psi])=\int D[n_y]\exp(-\mathcal{S}\left[\psi,n_y\right]). 
\end{equation}
The derivative of $\mathcal{S}_{ef}[\psi]$ with respect to $\psi$, lead us to:  
\begin{equation}
 \frac{\delta \mathcal{S}_{ef}[\psi]}{\delta \psi}=a\psi+b\psi^3+C_yk_0^2\langle n_y^2\rangle\psi,
 \label{eq:derS}
\end{equation}
where 
\begin{equation}
 \langle n_y^2\rangle=\int\frac{d^2kd\omega}{(2\pi)^3}\frac{1}{g_0\omega^2+gk^\sigma+C_yk_0^2\psi^2}.
\label{eq:ny2}
\end{equation}
Computing the integral in Eq. (\ref{eq:ny2}) at leading order in $\psi$ and replacing the result in Eq. (\ref{eq:derS}), we obtain: 
\begin{equation}
  \frac{\delta \mathcal{S}_{ef}[\psi]}{\delta \psi} = \left\{
  \begin{array}{lcl}
    \tilde{a}\psi-\tilde C_y k_0^2 \psi^{\frac{4}{\sigma}}+\tilde b\psi^3 & 
    \mbox{~,~} & \frac{4}{3}\leq\sigma\leq2\\
    & & \\
    \tilde{a}\psi+\tilde{b}\psi^3 & \mbox{~,~} & \sigma\leq\frac{4}{3}\\
  \end{array}\right. ,
\label{eq:Sderivative}
\end{equation}
where $\tilde a$, $\tilde b$ and $\tilde C_y$ are renormalized couplings.

The first line of Eq. (\ref{eq:Sderivative}) implies that the system has a first order smectic-nematic transition induced
by fluctuations for $\frac{4}{3}<\sigma\leq2$. This result includes the short-range interaction case $\sigma=2$. Moreover, 
it is clear from the second line of Eq. (\ref{eq:Sderivative}) that for  $\sigma\leq\frac{4}{3}$  the transition is  second order,
provided the nematic fluctuations are weak enough to keep  the renormalization of the quartic coupling $\tilde b$ positive. 
If this condition is not satisfied,  it is necessary to include higher order powers
of $\psi$ in the calculation. 

Summarizing, we expect that the two-dimensional smectic-nematic quantum  phase transition  is discontinuous for short range as
well as  long range interaction models, provided $\frac{4}{3}<\sigma\leq2$. The dipolar interaction is included in this regime. 
On the other hand, for extremely long ranged  interactions $\sigma\leq\frac{4}{3}$ (including the Coulomb case), the transition is 
of second order. The transitions are shown in figures \ref{fig:phasediag1} and \ref{fig:phasediag2}. It is interesting to note 
that there is no critical line ending at the quantum transition point since any finite temperature completely destroys positional
order. This fact could change in the presence of a substrate where the translation invariance is broken to a discrete group.
In this case, the smectic phase could be stabilized at finite temperature.

\section{Thermal Fluctuations}
\label{Sec:Thermal}
In order to understand the $T-r$ phase diagram, we analyze in this section thermal fluctuations. 
Similarly to the $T=0$ case, there are two very distinct behaviors depending on the range of interactions, which we
proceed to describe separately.

\subsection{$\sigma\ge 2$}
In this case, the model of eq. (\ref{eq:SoGL1})  coincides with the quantum rotor model in $d$ dimensions and finite $T$,  extensively discussed in the literature\cite{Sa2011}.
$d=2$ is a special case, because at finite $T$, it is the lower critical dimension. Thus, for negligible quantum fluctuations a Kosterlitz-Thoules (KT) classical phase transition takes place at a critical $T_{KT}$. Near the critical line, we expect that thermal fluctuations dominate (except at the  QCP). So, there should be a KT line that connects the classical and the quantum critical points as shown in figure \ref{fig:phasediag1}. The long range orientational order at $T=0$ is destroyed by temperature fluctuations, producing a quasi-long-range ordered (QLRO) phase, characterized by a power law decay of the order parameter correlation functions. This expectations are supported by   numerical simulations of the $XY$ model in three dimensions, with a finite third dimension~\cite{HoCa2014}. Although we do not expect that an $\epsilon=3-d$ expansion could produce sensible results for $d=2$, they will be useful for the long range interaction regime ($\sigma<2$) where  the lower critical dimension is $d=\sigma$.
In general, the inverse susceptibility very near the QCP is expected to be of the form:
\begin{equation}
\chi^{-1}(k,\delta r)=k^2+ R(\delta r,T),
\label{eq:chi2}
\end{equation}
where $R(\delta r,T)$ is the gap opened when one moves away from the QCP. $\delta r= r-r_c$ is the distance to the critical point.  This gap can be computed in the limits $\delta r/r\ll 1$ and  $T\gg\ \delta r^\nu$, leading to the following results\cite{Sa1997,Sa2011}:
\begin{equation}
R(\delta r,T)= \left\{
\begin{array}{lcl}
\delta r +c_1T^2 & \mbox{~for~} &  d\lesssim 3  \\
\delta r +c_1T^{d-1} & \mbox{~for~} &  d> 3
\end{array}
\right.
\label{eq:R2}
\end{equation}
where $c_1$ is a non-universal constant. 

\subsection{$\sigma <2$}
This long ranged interacting regime is very different from the previous one. The lower critical dimension is $d=\sigma$, then for $d=2$ we have a classical thermal second order phase transition for relatively small quantum fluctuations\cite{MeStNi2015}. Consequently, a second order line connects the classical and the quantum critical point, producing a whole region of the phase diagram of truly long ranged order, depicted in figure \ref{fig:phasediag2}.

To compute the low temperature behavior near the QCP we closely follow the methods of ref. \onlinecite{Sa1997}.
First, by summing up over the high frequency modes, we obtain an effective action for the zero frequency mode. The result is an effective ``classical'' theory whose parameters are renormalized by quantum fluctuations. Starting from the model of eq. (\ref{eq:SoGL1}) we obtain:
\begin{eqnarray}
\nonumber
\mathcal{S}_{T}&=&\frac{\beta}{2}\int \frac{d^dk}{(2\pi)^d}\ \left(k^\sigma+R\right)\hat{n}_\alpha(\vec{k})\cdot\hat{n}_\alpha(-\vec{k})\\ \nonumber
&+&\beta U\int\prod_{i=1}^4\frac{d^dk_i}{(2\pi)^d}\ \delta\left(\sum_{i=1}^4\vec{k}_i\right)\\ 
&\times&\hat{n}_\alpha(\vec{k}_1)\hat{n}_\alpha(\vec{k}_2)\hat{n}_\beta(\vec{k}_3)\hat{n}_\beta(\vec{k}_4),
\label{eq:ST}
\end{eqnarray}
where the couplings $R$ and $U$ are given by:
\begin{eqnarray}
 R(d,\sigma,r,u,\beta)&=&r+4u(N+2) \times \nonumber \\
 &\times& \frac{1}{\beta}\sum_{n\neq0}\int\frac{d^dk}{(2\pi)^d}\frac{1}{\omega_n^2+k^\sigma+r}, 
 \label{eq:Rbeta}
 \\ 
 U(d,\sigma,r,u,\beta)&=&u-4u^2(N+8)\times  \nonumber\\ 
 &\times&\frac{1}{\beta}\sum_{n\neq0}\int\frac{d^dk}{(2\pi)^d}\frac{1}{(\omega_n^2+k^\sigma+r)^2}.
 \label{eq:Ubeta}
\end{eqnarray}
From eq. (\ref{eq:ST}) it is straightforward to compute the inverse susceptibility at one loop approximation:
\begin{equation}
 \chi(k,d,\sigma,R,U)^{-1}=k^\sigma+R+2\frac{U}{4!}N\langle\hat{n}_{\alpha}^2\rangle,
 \label{eq:chibeta}
\end{equation}
where
\begin{equation}
\langle\hat{n}_{\alpha}^2\rangle=\frac{1}{\beta}\int\frac{d^dk}{(2\pi)^{d}}\frac{1}{k^\sigma+R+2\frac{U}{4!}N\langle\hat{n}_{\alpha}^2\rangle}.
\label{eq:nbeta}
\end{equation}
In these equations $R$ and $U$ are given by Eqs. (\ref{eq:Rbeta}) and (\ref{eq:Ubeta}) respectively.  

Before proceeding with the computation of the integrals it is interesting to  investigate how the susceptibility scales with the transformation (\ref{eq:relations}).
Form eqs. (\ref{eq:Rbeta}) and (\ref{eq:relations}) the gap parameter at 
$\sigma<2$ is related with the one computed at $\sigma=2$ as:
\begin{equation}
 R(d,\sigma,r,u,\beta)= R(d',2,r,u',\beta) \; .
\label{eq:RbetaScaling}
\end{equation}
Moreover, performing the same transformation on eq. (\ref{eq:Ubeta}) we find, 
\begin{equation}
U(d,\sigma,r,u,\beta)=\frac{2}{\sigma} f(d',\sigma)U(d',2,r,u',\beta)\; . 
\label{eq:UbetaScaling}
\end{equation}
Using these results we can show that, in the context of one loop approximation,  the susceptibility for $\sigma<2$, given by Eqs. (\ref{eq:chibeta}) and (\ref{eq:nbeta}),  is related with the one computed with   the short-ranged interaction model by the expression    
\begin{equation}
 \chi(k,d,\sigma,R,U,\beta)=\chi(k',d',2,R,U',\beta),
\label{eq:chibetaScaling}
\end{equation}
where the scaling relations are given by eq. (\ref{eq:relations}). Interestingly, the same scaling properties satisfied by the susceptibility   at $T=0$ are
 satisfied at finite, albeit small, temperatures.   

Equation (\ref{eq:chibetaScaling}) can be used to compute several critical properties of the long-ranged interaction model at finite temperature. The inverse susceptibility for $\sigma<2$ at finite temperature has a similar  expression as in the local case, 
\begin{equation}
\chi^{-1}(k,\delta r)=k^\sigma+ R(\delta r,T)\;.
\label{eq:chi}
\end{equation}
However, the gap parameter $R$ is now computed using the local model result (\ref{eq:R2}) and the scaling relations (\ref{eq:relations}). We find,  
\begin{equation}
R(\delta r,T)= \left\{
\begin{array}{lcl}
\delta r +c_1T^2 & \mbox{~for~} &  d< \frac{3\sigma}{2}  \\
\delta r +c_1T^{\frac{2d}{\sigma}-1} & \mbox{~for~} &  d> \frac{3\sigma}{2}
\end{array}
\right. \; .
\label{eq:R}
\end{equation}
The condition $R(\delta r(T),T)=0$ determines the critical line near the QCP:
\begin{equation}
\delta r(T)= \left\{
\begin{array}{lcl}
-c_1T^2 & \mbox{~for~} &  d< \frac{3\sigma}{2}  \\
-c_1T^{\frac{2d}{\sigma}-1} & \mbox{~for~} &  d> \frac{3\sigma}{2}
\end{array}
\right. \; .
\label{eq:r(T)}
\end{equation}
Moreover, the correlation length at the QCP ($\delta r=0$) as $T\to 0$ can be obtained from 
$\xi(T)=R(0,T)^{-{1}/{\sigma}}$, giving:
\begin{equation}
\xi(T)\sim \left\{
\begin{array}{lcl}
T^{-2/\sigma} & \mbox{~for~} &  d< \frac{3\sigma}{2}  \\
T^{-\frac{1}{\sigma}(\frac{2d}{\sigma}-1)} & \mbox{~for~} &  d> \frac{3\sigma}{2}.
\end{array}
\right.
\label{eq:xi(T)}
\end{equation}
These results are summarized in table \ref{tb:exponents} and figure \ref{fig:phasediag2}.

Let us conclude this section by discussing the validity of  the approximations made,  for the particular case $d=2$ with $\sigma\leq 2$.
In the usual  $O(N)$ model in $d_{\rm eff}=d+1$ the upper critical dimension is $d_u=3$ ($d_{\rm eff}=4$),  while the lower critical dimension is $d_l=2$. For this reason, figure \ref{fig:phasediag1} displays a KT line, there is no symmetry breaking across the critical line, and the low temperature phase is actually a QLRO phase. Evidently, an $\epsilon$ expansion around the upper critical dimension fails in this limit. However, when $\sigma<2$, the upper as well as the lower critical dimensions depends on $\sigma$ and they are actually given by $d_l=\sigma$, $d_u=3\sigma/2$.  Consequently, the two-dimensional system is always above the lower critical dimension. For this reason, the critical line depicted in figure \ref{fig:phasediag2}
is truly second order and an $\epsilon$ expansion produce  qualitatively good results. Of course, as usual,  the numerical values  computed at order $\epsilon$ have growing errors when we move away form the 
upper critical dimension $d_u=3\sigma/2$. It is interesting to note that for $d=2$ with Coulomb interactions ($\sigma=1$),  $d>d_u=3/2$ and then the critical exponents are dominated by the Gaussian fixed point and therefore are exact. The perturbative, finite $T$  calculations, provide the correlation length as well as the susceptibility diverging as $T^{-3}$ at the QCP,  while the critical line behaves as $\delta r\sim T^3$ while $T\to 0$.
  
\section{Discussion and Conclusions}
\label{Sec:Conclusions}
In this work we developed a theory for the melting of stripe phases in two dimensional quantum systems with competing interactions of
variable range, considering both quantum and thermal fluctuations. Our main conclusion is that the nature of the phase transitions as
a consequence of the melting process can be very different depending on the range of the competing repulsive interactions $\sigma$,
extending considerably the known results which, almost exclusively, were restricted to short range interactions. A mapping of the
problem to a model of quantum rotors in the plane with generalized dipolar interactions allowed us to obtain several interesting
properties of the phase transitions and universality classes of the models. At $T=0$ we showed that the melting of stripes proceeds
through a two step process, which can produce two quantum critical points for sufficiently long ranged repulsive interactions, $\sigma \leq 4/3$,
while when $\sigma > 4/3$ the smectic-nematic transition turns out to be of first order. At finite temperatures only some kind of
orientational order is possible. When $\sigma \geq 2$ the well known critical phase with algebraic orientational correlations is
present, ending at a KT line. But for sufficiently long range repulsive interactions $\sigma <2$ , a phase with long range nematic
order is possible, ending at a second order critical line. 
At $T=0$, critical exponents can be computed due the equivalence of the quantum $d=2$ problem at $T=0$ with the classical model of rotors
in $d+1=3$ dimensions at finite temperature, which properties for short range interactions are well known. An approximate treatment for finite temperatures allowed
us to compute the behavior of thermodynamic quantities near the QCP, specially the temperature dependence of the uniform susceptibility,
the correlation length and the critical line, summarized in Table \ref{tb:exponents}. 

In this work the melting of stripes defined by a scalar density order parameter was studied. 
Our results could be tested, e.g. in ultra-cold dipolar Fermi gases in the case where the dipoles point perpendicular
to the plane of the system, in which case the system of dipoles recovers space rotational invariance~\cite{Wu2016}. 
Important extensions for future work are the
consideration of a vector order parameter, e.g. inclusion of different spin components, and also an interaction among different degrees of
freedom, which is important for the physics of high $T_c$ superconductors and other strongly correlated electronic systems in which 
electron and spin density waves are intertwined, as considered, e.g. in [\onlinecite{Senthil2012}]. 
Lattice anisotropies can be also important perturbations in real situations which deserve to be studied in future work. 

\acknowledgments
The Brazilian agencies {\em Conselho Nacional de Desenvolvimento Cient\'\i fico e Tecnol\'ogico} (CNPq), {\em Funda\c c\~ao  Carlos Chagas Filho de Amparo \`a Pesquisa do Estado do Rio de Janeiro} (FAPERJ), and {\em Coordena\c c\~ao de Aperfei\c coamento de Pessoal de N\'\i vel Superior} (CAPES) are acknowledged for partial financial support.  D.G.B also acknowledges partial financial support by the  Associate Program of the Abdus Salam International Centre for Theoretical Physics, ICTP, Trieste, Italy.


\begin{thebibliography}{62}%
\makeatletter
\providecommand \@ifxundefined [1]{%
 \@ifx{#1\undefined}
}%
\providecommand \@ifnum [1]{%
 \ifnum #1\expandafter \@firstoftwo
 \else \expandafter \@secondoftwo
 \fi
}%
\providecommand \@ifx [1]{%
 \ifx #1\expandafter \@firstoftwo
 \else \expandafter \@secondoftwo
 \fi
}%
\providecommand \natexlab [1]{#1}%
\providecommand \enquote  [1]{``#1''}%
\providecommand \bibnamefont  [1]{#1}%
\providecommand \bibfnamefont [1]{#1}%
\providecommand \citenamefont [1]{#1}%
\providecommand \href@noop [0]{\@secondoftwo}%
\providecommand \href [0]{\begingroup \@sanitize@url \@href}%
\providecommand \@href[1]{\@@startlink{#1}\@@href}%
\providecommand \@@href[1]{\endgroup#1\@@endlink}%
\providecommand \@sanitize@url [0]{\catcode `\\12\catcode `\$12\catcode
  `\&12\catcode `\#12\catcode `\^12\catcode `\_12\catcode `\%12\relax}%
\providecommand \@@startlink[1]{}%
\providecommand \@@endlink[0]{}%
\providecommand \url  [0]{\begingroup\@sanitize@url \@url }%
\providecommand \@url [1]{\endgroup\@href {#1}{\urlprefix }}%
\providecommand \urlprefix  [0]{URL }%
\providecommand \Eprint [0]{\href }%
\providecommand \doibase [0]{http://dx.doi.org/}%
\providecommand \selectlanguage [0]{\@gobble}%
\providecommand \bibinfo  [0]{\@secondoftwo}%
\providecommand \bibfield  [0]{\@secondoftwo}%
\providecommand \translation [1]{[#1]}%
\providecommand \BibitemOpen [0]{}%
\providecommand \bibitemStop [0]{}%
\providecommand \bibitemNoStop [0]{.\EOS\space}%
\providecommand \EOS [0]{\spacefactor3000\relax}%
\providecommand \BibitemShut  [1]{\csname bibitem#1\endcsname}%
\let\auto@bib@innerbib\@empty
\bibitem [{\citenamefont {Kivelson}\ \emph {et~al.}(2003)\citenamefont
  {Kivelson}, \citenamefont {Bindloss}, \citenamefont {Fradkin}, \citenamefont
  {Oganesyan}, \citenamefont {Tranquada}, \citenamefont {Kapitulnik},\ and\
  \citenamefont {Howald}}]{KivFrad2003}%
  \BibitemOpen
  \bibfield  {author} {\bibinfo {author} {\bibfnamefont {S.~A.}\ \bibnamefont
  {Kivelson}}, \bibinfo {author} {\bibfnamefont {I.~P.}\ \bibnamefont
  {Bindloss}}, \bibinfo {author} {\bibfnamefont {E.}~\bibnamefont {Fradkin}},
  \bibinfo {author} {\bibfnamefont {V.}~\bibnamefont {Oganesyan}}, \bibinfo
  {author} {\bibfnamefont {J.~M.}\ \bibnamefont {Tranquada}}, \bibinfo {author}
  {\bibfnamefont {A.}~\bibnamefont {Kapitulnik}}, \ and\ \bibinfo {author}
  {\bibfnamefont {C.}~\bibnamefont {Howald}},\ }\href {\doibase
  10.1103/RevModPhys.75.1201} {\bibfield  {journal} {\bibinfo  {journal} {Rev.
  Mod. Phys.}\ }\textbf {\bibinfo {volume} {75}},\ \bibinfo {pages} {1201}
  (\bibinfo {year} {2003})}\BibitemShut {NoStop}%
\bibitem [{\citenamefont {Vojta}(2009)}]{Vojta2009}%
  \BibitemOpen
  \bibfield  {author} {\bibinfo {author} {\bibfnamefont {M.}~\bibnamefont
  {Vojta}},\ }\href {\doibase 10.1080/00018730903122242} {\bibfield  {journal}
  {\bibinfo  {journal} {Advances in Physics}\ }\textbf {\bibinfo {volume}
  {58}},\ \bibinfo {pages} {699} (\bibinfo {year} {2009})}\BibitemShut
  {NoStop}%
\bibitem [{\citenamefont {Fradkin}\ \emph {et~al.}(2015)\citenamefont
  {Fradkin}, \citenamefont {Kivelson},\ and\ \citenamefont
  {Tranquada}}]{FrKiTr2015}%
  \BibitemOpen
  \bibfield  {author} {\bibinfo {author} {\bibfnamefont {E.}~\bibnamefont
  {Fradkin}}, \bibinfo {author} {\bibfnamefont {S.~A.}\ \bibnamefont
  {Kivelson}}, \ and\ \bibinfo {author} {\bibfnamefont {J.~M.}\ \bibnamefont
  {Tranquada}},\ }\href {\doibase 10.1103/RevModPhys.87.457} {\bibfield
  {journal} {\bibinfo  {journal} {Rev. Mod. Phys.}\ }\textbf {\bibinfo {volume}
  {87}},\ \bibinfo {pages} {457} (\bibinfo {year} {2015})}\BibitemShut
  {NoStop}%
\bibitem [{\citenamefont {Fogler}\ \emph {et~al.}(1996)\citenamefont {Fogler},
  \citenamefont {Koulakov},\ and\ \citenamefont {Shklovskii}}]{Fogler1996}%
  \BibitemOpen
  \bibfield  {author} {\bibinfo {author} {\bibfnamefont {M.~M.}\ \bibnamefont
  {Fogler}}, \bibinfo {author} {\bibfnamefont {A.~A.}\ \bibnamefont
  {Koulakov}}, \ and\ \bibinfo {author} {\bibfnamefont {B.~I.}\ \bibnamefont
  {Shklovskii}},\ }\href {\doibase 10.1103/PhysRevB.54.1853} {\bibfield
  {journal} {\bibinfo  {journal} {Phys. Rev. B}\ }\textbf {\bibinfo {volume}
  {54}},\ \bibinfo {pages} {1853} (\bibinfo {year} {1996})}\BibitemShut
  {NoStop}%
\bibitem [{\citenamefont {Lilly}\ \emph {et~al.}(1999)\citenamefont {Lilly},
  \citenamefont {Cooper}, \citenamefont {Eisenstein}, \citenamefont
  {Pfeiffer},\ and\ \citenamefont {West}}]{Lilly1999}%
  \BibitemOpen
  \bibfield  {author} {\bibinfo {author} {\bibfnamefont {M.~P.}\ \bibnamefont
  {Lilly}}, \bibinfo {author} {\bibfnamefont {K.~B.}\ \bibnamefont {Cooper}},
  \bibinfo {author} {\bibfnamefont {J.~P.}\ \bibnamefont {Eisenstein}},
  \bibinfo {author} {\bibfnamefont {L.~N.}\ \bibnamefont {Pfeiffer}}, \ and\
  \bibinfo {author} {\bibfnamefont {K.~W.}\ \bibnamefont {West}},\ }\href
  {\doibase 10.1103/PhysRevLett.82.394} {\bibfield  {journal} {\bibinfo
  {journal} {Phys. Rev. Lett.}\ }\textbf {\bibinfo {volume} {82}},\ \bibinfo
  {pages} {394} (\bibinfo {year} {1999})}\BibitemShut {NoStop}%
\bibitem [{\citenamefont {Fradkin}\ and\ \citenamefont
  {Kivelson}(1999)}]{FrKi1999}%
  \BibitemOpen
  \bibfield  {author} {\bibinfo {author} {\bibfnamefont {E.}~\bibnamefont
  {Fradkin}}\ and\ \bibinfo {author} {\bibfnamefont {S.~A.}\ \bibnamefont
  {Kivelson}},\ }\href {\doibase 10.1103/PhysRevB.59.8065} {\bibfield
  {journal} {\bibinfo  {journal} {Phys. Rev. B}\ }\textbf {\bibinfo {volume}
  {59}},\ \bibinfo {pages} {8065} (\bibinfo {year} {1999})}\BibitemShut
  {NoStop}%
\bibitem [{\citenamefont {Friess}\ \emph {et~al.}(2014)\citenamefont {Friess},
  \citenamefont {Umansky}, \citenamefont {Tiemann}, \citenamefont {von
  Klitzing},\ and\ \citenamefont {Smet}}]{Friess2014}%
  \BibitemOpen
  \bibfield  {author} {\bibinfo {author} {\bibfnamefont {B.}~\bibnamefont
  {Friess}}, \bibinfo {author} {\bibfnamefont {V.}~\bibnamefont {Umansky}},
  \bibinfo {author} {\bibfnamefont {L.}~\bibnamefont {Tiemann}}, \bibinfo
  {author} {\bibfnamefont {K.}~\bibnamefont {von Klitzing}}, \ and\ \bibinfo
  {author} {\bibfnamefont {J.~H.}\ \bibnamefont {Smet}},\ }\href {\doibase
  10.1103/PhysRevLett.113.076803} {\bibfield  {journal} {\bibinfo  {journal}
  {Phys. Rev. Lett.}\ }\textbf {\bibinfo {volume} {113}},\ \bibinfo {pages}
  {076803} (\bibinfo {year} {2014})}\BibitemShut {NoStop}%
\bibitem [{\citenamefont {Parker}\ \emph {et~al.}(2010)\citenamefont {Parker},
  \citenamefont {Aynajian}, \citenamefont {da~Silva~Neto}, \citenamefont
  {Pushp}, \citenamefont {Ono}, \citenamefont {Wen}, \citenamefont {Xu},
  \citenamefont {Gu},\ and\ \citenamefont {Yazdani}}]{Parker2010}%
  \BibitemOpen
  \bibfield  {author} {\bibinfo {author} {\bibfnamefont {C.~V.}\ \bibnamefont
  {Parker}}, \bibinfo {author} {\bibfnamefont {P.}~\bibnamefont {Aynajian}},
  \bibinfo {author} {\bibfnamefont {E.~H.}\ \bibnamefont {da~Silva~Neto}},
  \bibinfo {author} {\bibfnamefont {A.}~\bibnamefont {Pushp}}, \bibinfo
  {author} {\bibfnamefont {S.}~\bibnamefont {Ono}}, \bibinfo {author}
  {\bibfnamefont {J.}~\bibnamefont {Wen}}, \bibinfo {author} {\bibfnamefont
  {Z.}~\bibnamefont {Xu}}, \bibinfo {author} {\bibfnamefont {G.}~\bibnamefont
  {Gu}}, \ and\ \bibinfo {author} {\bibfnamefont {A.}~\bibnamefont {Yazdani}},\
  }\href {http://dx.doi.org/10.1038/nature09597} {\bibfield  {journal}
  {\bibinfo  {journal} {Nature}\ }\textbf {\bibinfo {volume} {468}},\ \bibinfo
  {pages} {677} (\bibinfo {year} {2010})}\BibitemShut {NoStop}%
\bibitem [{\citenamefont {Daou}\ \emph {et~al.}(2010)\citenamefont {Daou},
  \citenamefont {Chang}, \citenamefont {LeBoeuf}, \citenamefont
  {Cyr-Choiniere}, \citenamefont {Laliberte}, \citenamefont {Doiron-Leyraud},
  \citenamefont {Ramshaw}, \citenamefont {Liang}, \citenamefont {Bonn},
  \citenamefont {Hardy},\ and\ \citenamefont {Taillefer}}]{Daou2010}%
  \BibitemOpen
  \bibfield  {author} {\bibinfo {author} {\bibfnamefont {R.}~\bibnamefont
  {Daou}}, \bibinfo {author} {\bibfnamefont {J.}~\bibnamefont {Chang}},
  \bibinfo {author} {\bibfnamefont {D.}~\bibnamefont {LeBoeuf}}, \bibinfo
  {author} {\bibfnamefont {O.}~\bibnamefont {Cyr-Choiniere}}, \bibinfo {author}
  {\bibfnamefont {F.}~\bibnamefont {Laliberte}}, \bibinfo {author}
  {\bibfnamefont {N.}~\bibnamefont {Doiron-Leyraud}}, \bibinfo {author}
  {\bibfnamefont {B.~J.}\ \bibnamefont {Ramshaw}}, \bibinfo {author}
  {\bibfnamefont {R.}~\bibnamefont {Liang}}, \bibinfo {author} {\bibfnamefont
  {D.~A.}\ \bibnamefont {Bonn}}, \bibinfo {author} {\bibfnamefont {W.~N.}\
  \bibnamefont {Hardy}}, \ and\ \bibinfo {author} {\bibfnamefont
  {L.}~\bibnamefont {Taillefer}},\ }\href
  {http://dx.doi.org/10.1038/nature08716} {\bibfield  {journal} {\bibinfo
  {journal} {Nature}\ }\textbf {\bibinfo {volume} {463}},\ \bibinfo {pages}
  {519} (\bibinfo {year} {2010})}\BibitemShut {NoStop}%
\bibitem [{\citenamefont {Tanatar}\ \emph {et~al.}(2016)\citenamefont
  {Tanatar}, \citenamefont {B\"ohmer}, \citenamefont {Timmons}, \citenamefont
  {Sch\"utt}, \citenamefont {Drachuck}, \citenamefont {Taufour}, \citenamefont
  {Kothapalli}, \citenamefont {Kreyssig}, \citenamefont {Bud'ko}, \citenamefont
  {Canfield}, \citenamefont {Fernandes},\ and\ \citenamefont
  {Prozorov}}]{Tanatar2016}%
  \BibitemOpen
  \bibfield  {author} {\bibinfo {author} {\bibfnamefont {M.~A.}\ \bibnamefont
  {Tanatar}}, \bibinfo {author} {\bibfnamefont {A.~E.}\ \bibnamefont
  {B\"ohmer}}, \bibinfo {author} {\bibfnamefont {E.~I.}\ \bibnamefont
  {Timmons}}, \bibinfo {author} {\bibfnamefont {M.}~\bibnamefont {Sch\"utt}},
  \bibinfo {author} {\bibfnamefont {G.}~\bibnamefont {Drachuck}}, \bibinfo
  {author} {\bibfnamefont {V.}~\bibnamefont {Taufour}}, \bibinfo {author}
  {\bibfnamefont {K.}~\bibnamefont {Kothapalli}}, \bibinfo {author}
  {\bibfnamefont {A.}~\bibnamefont {Kreyssig}}, \bibinfo {author}
  {\bibfnamefont {S.~L.}\ \bibnamefont {Bud'ko}}, \bibinfo {author}
  {\bibfnamefont {P.~C.}\ \bibnamefont {Canfield}}, \bibinfo {author}
  {\bibfnamefont {R.~M.}\ \bibnamefont {Fernandes}}, \ and\ \bibinfo {author}
  {\bibfnamefont {R.}~\bibnamefont {Prozorov}},\ }\href {\doibase
  10.1103/PhysRevLett.117.127001} {\bibfield  {journal} {\bibinfo  {journal}
  {Phys. Rev. Lett.}\ }\textbf {\bibinfo {volume} {117}},\ \bibinfo {pages}
  {127001} (\bibinfo {year} {2016})}\BibitemShut {NoStop}%
\bibitem [{\citenamefont {Grigera}\ \emph {et~al.}(2001)\citenamefont
  {Grigera}, \citenamefont {Perry}, \citenamefont {Schofield}, \citenamefont
  {Chiao}, \citenamefont {Julian}, \citenamefont {Lonzarich}, \citenamefont
  {Ikeda}, \citenamefont {Maeno}, \citenamefont {Millis},\ and\ \citenamefont
  {Mackenzie}}]{Grigera2001}%
  \BibitemOpen
  \bibfield  {author} {\bibinfo {author} {\bibfnamefont {S.~A.}\ \bibnamefont
  {Grigera}}, \bibinfo {author} {\bibfnamefont {R.~S.}\ \bibnamefont {Perry}},
  \bibinfo {author} {\bibfnamefont {A.~J.}\ \bibnamefont {Schofield}}, \bibinfo
  {author} {\bibfnamefont {M.}~\bibnamefont {Chiao}}, \bibinfo {author}
  {\bibfnamefont {S.~R.}\ \bibnamefont {Julian}}, \bibinfo {author}
  {\bibfnamefont {G.~G.}\ \bibnamefont {Lonzarich}}, \bibinfo {author}
  {\bibfnamefont {S.~I.}\ \bibnamefont {Ikeda}}, \bibinfo {author}
  {\bibfnamefont {Y.}~\bibnamefont {Maeno}}, \bibinfo {author} {\bibfnamefont
  {A.~J.}\ \bibnamefont {Millis}}, \ and\ \bibinfo {author} {\bibfnamefont
  {A.~P.}\ \bibnamefont {Mackenzie}},\ }\href {\doibase
  10.1126/science.1063539} {\bibfield  {journal} {\bibinfo  {journal}
  {Science}\ }\textbf {\bibinfo {volume} {294}},\ \bibinfo {pages} {329}
  (\bibinfo {year} {2001})},\ \Eprint
  {http://arxiv.org/abs/http://www.sciencemag.org/content/294/5541/329.full.pdf}
  {http://www.sciencemag.org/content/294/5541/329.full.pdf} \BibitemShut
  {NoStop}%
\bibitem [{\citenamefont {Borzi}\ \emph {et~al.}(2007)\citenamefont {Borzi},
  \citenamefont {Grigera}, \citenamefont {Farrell}, \citenamefont {Perry},
  \citenamefont {Lister}, \citenamefont {Lee}, \citenamefont {Tennant},
  \citenamefont {Maeno},\ and\ \citenamefont {Mackenzie}}]{Borzi2007}%
  \BibitemOpen
  \bibfield  {author} {\bibinfo {author} {\bibfnamefont {R.~A.}\ \bibnamefont
  {Borzi}}, \bibinfo {author} {\bibfnamefont {S.~A.}\ \bibnamefont {Grigera}},
  \bibinfo {author} {\bibfnamefont {J.}~\bibnamefont {Farrell}}, \bibinfo
  {author} {\bibfnamefont {R.~S.}\ \bibnamefont {Perry}}, \bibinfo {author}
  {\bibfnamefont {S.~J.~S.}\ \bibnamefont {Lister}}, \bibinfo {author}
  {\bibfnamefont {S.~L.}\ \bibnamefont {Lee}}, \bibinfo {author} {\bibfnamefont
  {D.~A.}\ \bibnamefont {Tennant}}, \bibinfo {author} {\bibfnamefont
  {Y.}~\bibnamefont {Maeno}}, \ and\ \bibinfo {author} {\bibfnamefont {A.~P.}\
  \bibnamefont {Mackenzie}},\ }\href {\doibase 10.1126/science.1134796}
  {\bibfield  {journal} {\bibinfo  {journal} {Science}\ }\textbf {\bibinfo
  {volume} {315}},\ \bibinfo {pages} {214} (\bibinfo {year} {2007})},\ \Eprint
  {http://arxiv.org/abs/http://www.sciencemag.org/content/315/5809/214.full.pdf}
  {http://www.sciencemag.org/content/315/5809/214.full.pdf} \BibitemShut
  {NoStop}%
\bibitem [{\citenamefont {Taillefer}(2010)}]{Taillefer2010}%
  \BibitemOpen
  \bibfield  {author} {\bibinfo {author} {\bibfnamefont {L.}~\bibnamefont
  {Taillefer}},\ }\href {\doibase 10.1146/annurev-conmatphys-070909-104117}
  {\bibfield  {journal} {\bibinfo  {journal} {Annual Reviews of Condensed
  Matter Physics}\ }\textbf {\bibinfo {volume} {1}},\ \bibinfo {pages} {51}
  (\bibinfo {year} {2010})},\ \Eprint
  {http://arxiv.org/abs/http://dx.doi.org/10.1146/annurev-conmatphys-070909-104117}
  {http://dx.doi.org/10.1146/annurev-conmatphys-070909-104117} \BibitemShut
  {NoStop}%
\bibitem [{\citenamefont {Toner}\ and\ \citenamefont
  {Nelson}(1981)}]{ToNe1981}%
  \BibitemOpen
  \bibfield  {author} {\bibinfo {author} {\bibfnamefont {J.}~\bibnamefont
  {Toner}}\ and\ \bibinfo {author} {\bibfnamefont {D.~R.}\ \bibnamefont
  {Nelson}},\ }\href {\doibase 10.1103/PhysRevB.23.316} {\bibfield  {journal}
  {\bibinfo  {journal} {Phys. Rev. B}\ }\textbf {\bibinfo {volume} {23}},\
  \bibinfo {pages} {316} (\bibinfo {year} {1981})}\BibitemShut {NoStop}%
\bibitem [{\citenamefont {Abanov}\ \emph {et~al.}(1995)\citenamefont {Abanov},
  \citenamefont {Kalatsky}, \citenamefont {Pokrovsky},\ and\ \citenamefont
  {Saslow}}]{AbKaPoSa1995}%
  \BibitemOpen
  \bibfield  {author} {\bibinfo {author} {\bibfnamefont {A.}~\bibnamefont
  {Abanov}}, \bibinfo {author} {\bibfnamefont {V.}~\bibnamefont {Kalatsky}},
  \bibinfo {author} {\bibfnamefont {V.~L.}\ \bibnamefont {Pokrovsky}}, \ and\
  \bibinfo {author} {\bibfnamefont {W.~M.}\ \bibnamefont {Saslow}},\
  }\href@noop {} {\bibfield  {journal} {\bibinfo  {journal} {Phys. Rev. B}\
  }\textbf {\bibinfo {volume} {51}},\ \bibinfo {pages} {1023} (\bibinfo {year}
  {1995})}\BibitemShut {NoStop}%
\bibitem [{\citenamefont {Wexler}\ and\ \citenamefont
  {Dorsey}(2001)}]{Wexler2001}%
  \BibitemOpen
  \bibfield  {author} {\bibinfo {author} {\bibfnamefont {C.}~\bibnamefont
  {Wexler}}\ and\ \bibinfo {author} {\bibfnamefont {A.~T.}\ \bibnamefont
  {Dorsey}},\ }\href {\doibase 10.1103/PhysRevB.64.115312} {\bibfield
  {journal} {\bibinfo  {journal} {Phys. Rev. B}\ }\textbf {\bibinfo {volume}
  {64}},\ \bibinfo {pages} {115312} (\bibinfo {year} {2001})}\BibitemShut
  {NoStop}%
\bibitem [{\citenamefont {Sachdev}(1997)}]{Sa1997}%
  \BibitemOpen
  \bibfield  {author} {\bibinfo {author} {\bibfnamefont {S.}~\bibnamefont
  {Sachdev}},\ }\href {\doibase 10.1103/PhysRevB.55.142} {\bibfield  {journal}
  {\bibinfo  {journal} {Phys. Rev. B}\ }\textbf {\bibinfo {volume} {55}},\
  \bibinfo {pages} {142} (\bibinfo {year} {1997})}\BibitemShut {NoStop}%
\bibitem [{\citenamefont {Continentino}(2001)}]{Mucio2001}%
  \BibitemOpen
  \bibfield  {author} {\bibinfo {author} {\bibfnamefont {M.~A.}\ \bibnamefont
  {Continentino}},\ }\href
  {http://gen.lib.rus.ec/book/index.php?md5=D95C8BA24F70B2291C7A02CDF2039E91}
  {\emph {\bibinfo {title} {Quantum scaling in many-body systems}}},\ \bibinfo
  {edition} {1st}\ ed.,\ World Scientific lecture notes in physics 67\
  (\bibinfo  {publisher} {World Scientific},\ \bibinfo {year}
  {2001})\BibitemShut {NoStop}%
\bibitem [{\citenamefont {Sachdev}(2011)}]{Sa2011}%
  \BibitemOpen
  \bibfield  {author} {\bibinfo {author} {\bibfnamefont {S.}~\bibnamefont
  {Sachdev}},\ }\href@noop {} {\emph {\bibinfo {title} {Quantum Phase
  Transitions}}}\ (\bibinfo  {publisher} {Cambridge University Press},\
  \bibinfo {year} {2011})\BibitemShut {NoStop}%
\bibitem [{\citenamefont {Mross}\ and\ \citenamefont
  {Senthil}(2012)}]{Senthil2012}%
  \BibitemOpen
  \bibfield  {author} {\bibinfo {author} {\bibfnamefont {D.~F.}\ \bibnamefont
  {Mross}}\ and\ \bibinfo {author} {\bibfnamefont {T.}~\bibnamefont
  {Senthil}},\ }\href {\doibase 10.1103/PhysRevB.86.115138} {\bibfield
  {journal} {\bibinfo  {journal} {Phys. Rev. B}\ }\textbf {\bibinfo {volume}
  {86}},\ \bibinfo {pages} {115138} (\bibinfo {year} {2012})}\BibitemShut
  {NoStop}%
\bibitem [{\citenamefont {Cho}\ \emph {et~al.}(2015)\citenamefont {Cho},
  \citenamefont {Parrikar}, \citenamefont {You}, \citenamefont {Leigh},\ and\
  \citenamefont {Hughes}}]{Cho2015}%
  \BibitemOpen
  \bibfield  {author} {\bibinfo {author} {\bibfnamefont {G.~Y.}\ \bibnamefont
  {Cho}}, \bibinfo {author} {\bibfnamefont {O.}~\bibnamefont {Parrikar}},
  \bibinfo {author} {\bibfnamefont {Y.}~\bibnamefont {You}}, \bibinfo {author}
  {\bibfnamefont {R.~G.}\ \bibnamefont {Leigh}}, \ and\ \bibinfo {author}
  {\bibfnamefont {T.~L.}\ \bibnamefont {Hughes}},\ }\href {\doibase
  10.1103/PhysRevB.91.035122} {\bibfield  {journal} {\bibinfo  {journal} {Phys.
  Rev. B}\ }\textbf {\bibinfo {volume} {91}},\ \bibinfo {pages} {035122}
  (\bibinfo {year} {2015})}\BibitemShut {NoStop}%
\bibitem [{\citenamefont {Beekman}\ \emph {et~al.}(2016)\citenamefont
  {Beekman}, \citenamefont {Nissinen}, \citenamefont {Kai~Wu}, \citenamefont
  {Slager}, \citenamefont {Nussinov}, \citenamefont {Cvetkovic},\ and\
  \citenamefont {Zaanen}}]{Beekman2016}%
  \BibitemOpen
  \bibfield  {author} {\bibinfo {author} {\bibfnamefont {A.~J.}\ \bibnamefont
  {Beekman}}, \bibinfo {author} {\bibfnamefont {J.}~\bibnamefont {Nissinen}},
  \bibinfo {author} {\bibfnamefont {K.~L.}\ \bibnamefont {Kai~Wu}}, \bibinfo
  {author} {\bibfnamefont {R.-J.}\ \bibnamefont {Slager}}, \bibinfo {author}
  {\bibfnamefont {Z.}~\bibnamefont {Nussinov}}, \bibinfo {author}
  {\bibfnamefont {V.}~\bibnamefont {Cvetkovic}}, \ and\ \bibinfo {author}
  {\bibfnamefont {J.}~\bibnamefont {Zaanen}},\ }\href@noop {} {\bibfield
  {journal} {\bibinfo  {journal} {arXiv}\ }\textbf {\bibinfo {volume}
  {1603.04254}} (\bibinfo {year} {2016})}\BibitemShut {NoStop}%
\bibitem [{\citenamefont {Radzihovsky}(2012)}]{Radzihovsky-2012}%
  \BibitemOpen
  \bibfield  {author} {\bibinfo {author} {\bibfnamefont {L.}~\bibnamefont
  {Radzihovsky}},\ }\href {\doibase
  http://dx.doi.org/10.1016/j.physc.2012.04.014} {\bibfield  {journal}
  {\bibinfo  {journal} {Physica C: Superconductivity}\ }\textbf {\bibinfo
  {volume} {481}},\ \bibinfo {pages} {189 } (\bibinfo {year}
  {2012})}\BibitemShut {NoStop}%
\bibitem [{\citenamefont {Lu}\ \emph {et~al.}(2012)\citenamefont {Lu},
  \citenamefont {Burdick},\ and\ \citenamefont {Lev}}]{Lu2012}%
  \BibitemOpen
  \bibfield  {author} {\bibinfo {author} {\bibfnamefont {M.}~\bibnamefont
  {Lu}}, \bibinfo {author} {\bibfnamefont {N.~Q.}\ \bibnamefont {Burdick}}, \
  and\ \bibinfo {author} {\bibfnamefont {B.~L.}\ \bibnamefont {Lev}},\ }\href
  {\doibase 10.1103/PhysRevLett.108.215301} {\bibfield  {journal} {\bibinfo
  {journal} {Phys. Rev. Lett.}\ }\textbf {\bibinfo {volume} {108}},\ \bibinfo
  {pages} {215301} (\bibinfo {year} {2012})}\BibitemShut {NoStop}%
\bibitem [{\citenamefont {Park}\ \emph {et~al.}(2015)\citenamefont {Park},
  \citenamefont {Will},\ and\ \citenamefont {Zwierlein}}]{Park2015}%
  \BibitemOpen
  \bibfield  {author} {\bibinfo {author} {\bibfnamefont {J.~W.}\ \bibnamefont
  {Park}}, \bibinfo {author} {\bibfnamefont {S.~A.}\ \bibnamefont {Will}}, \
  and\ \bibinfo {author} {\bibfnamefont {M.~W.}\ \bibnamefont {Zwierlein}},\
  }\href {\doibase 10.1103/PhysRevLett.114.205302} {\bibfield  {journal}
  {\bibinfo  {journal} {Phys. Rev. Lett.}\ }\textbf {\bibinfo {volume} {114}},\
  \bibinfo {pages} {205302} (\bibinfo {year} {2015})}\BibitemShut {NoStop}%
\bibitem [{\citenamefont {Yamaguchi}\ \emph {et~al.}(2010)\citenamefont
  {Yamaguchi}, \citenamefont {Sogo}, \citenamefont {Ito},\ and\ \citenamefont
  {Miyakawa}}]{Yamaguchi2010}%
  \BibitemOpen
  \bibfield  {author} {\bibinfo {author} {\bibfnamefont {Y.}~\bibnamefont
  {Yamaguchi}}, \bibinfo {author} {\bibfnamefont {T.}~\bibnamefont {Sogo}},
  \bibinfo {author} {\bibfnamefont {T.}~\bibnamefont {Ito}}, \ and\ \bibinfo
  {author} {\bibfnamefont {T.}~\bibnamefont {Miyakawa}},\ }\href {\doibase
  10.1103/PhysRevA.82.013643} {\bibfield  {journal} {\bibinfo  {journal} {Phys.
  Rev. A}\ }\textbf {\bibinfo {volume} {82}},\ \bibinfo {pages} {013643}
  (\bibinfo {year} {2010})}\BibitemShut {NoStop}%
\bibitem [{\citenamefont {Gorshkov}\ \emph {et~al.}(2011)\citenamefont
  {Gorshkov}, \citenamefont {Manmana}, \citenamefont {Chen}, \citenamefont
  {Ye}, \citenamefont {Demler}, \citenamefont {Lukin},\ and\ \citenamefont
  {Rey}}]{Gorshkov2011}%
  \BibitemOpen
  \bibfield  {author} {\bibinfo {author} {\bibfnamefont {A.~V.}\ \bibnamefont
  {Gorshkov}}, \bibinfo {author} {\bibfnamefont {S.~R.}\ \bibnamefont
  {Manmana}}, \bibinfo {author} {\bibfnamefont {G.}~\bibnamefont {Chen}},
  \bibinfo {author} {\bibfnamefont {J.}~\bibnamefont {Ye}}, \bibinfo {author}
  {\bibfnamefont {E.}~\bibnamefont {Demler}}, \bibinfo {author} {\bibfnamefont
  {M.~D.}\ \bibnamefont {Lukin}}, \ and\ \bibinfo {author} {\bibfnamefont
  {A.~M.}\ \bibnamefont {Rey}},\ }\href {\doibase
  10.1103/PhysRevLett.107.115301} {\bibfield  {journal} {\bibinfo  {journal}
  {Phys. Rev. Lett.}\ }\textbf {\bibinfo {volume} {107}},\ \bibinfo {pages}
  {115301} (\bibinfo {year} {2011})}\BibitemShut {NoStop}%
\bibitem [{\citenamefont {Babadi}\ and\ \citenamefont
  {Demler}(2011)}]{Babadi2011}%
  \BibitemOpen
  \bibfield  {author} {\bibinfo {author} {\bibfnamefont {M.}~\bibnamefont
  {Babadi}}\ and\ \bibinfo {author} {\bibfnamefont {E.}~\bibnamefont
  {Demler}},\ }\href {\doibase 10.1103/PhysRevB.84.235124} {\bibfield
  {journal} {\bibinfo  {journal} {Phys. Rev. B}\ }\textbf {\bibinfo {volume}
  {84}},\ \bibinfo {pages} {235124} (\bibinfo {year} {2011})}\BibitemShut
  {NoStop}%
\bibitem [{\citenamefont {Parish}\ and\ \citenamefont
  {Marchetti}(2012)}]{Parish2012}%
  \BibitemOpen
  \bibfield  {author} {\bibinfo {author} {\bibfnamefont {M.~M.}\ \bibnamefont
  {Parish}}\ and\ \bibinfo {author} {\bibfnamefont {F.~M.}\ \bibnamefont
  {Marchetti}},\ }\href {\doibase 10.1103/PhysRevLett.108.145304} {\bibfield
  {journal} {\bibinfo  {journal} {Phys. Rev. Lett.}\ }\textbf {\bibinfo
  {volume} {108}},\ \bibinfo {pages} {145304} (\bibinfo {year}
  {2012})}\BibitemShut {NoStop}%
\bibitem [{\citenamefont {Wu}\ \emph {et~al.}()\citenamefont {Wu},
  \citenamefont {Block},\ and\ \citenamefont {Bruun}}]{Wu2016}%
  \BibitemOpen
  \bibfield  {author} {\bibinfo {author} {\bibfnamefont {Z.}~\bibnamefont
  {Wu}}, \bibinfo {author} {\bibfnamefont {J.~K.}\ \bibnamefont {Block}}, \
  and\ \bibinfo {author} {\bibfnamefont {G.~M.}\ \bibnamefont {Bruun}},\
  }\href@noop {} {\bibfield  {journal} {\bibinfo  {journal} {Scientific
  Reports}\ }\textbf {\bibinfo {volume} {6}},\ \bibinfo {pages}
  {19038}}\BibitemShut {NoStop}%
\bibitem [{\citenamefont {Barci}\ and\ \citenamefont
  {Stariolo}(2009)}]{BaSt2009}%
  \BibitemOpen
  \bibfield  {author} {\bibinfo {author} {\bibfnamefont {D.~G.}\ \bibnamefont
  {Barci}}\ and\ \bibinfo {author} {\bibfnamefont {D.~A.}\ \bibnamefont
  {Stariolo}},\ }\href {\doibase 10.1103/PhysRevB.79.075437} {\bibfield
  {journal} {\bibinfo  {journal} {Physical Review B}\ }\textbf {\bibinfo
  {volume} {79}},\ \bibinfo {eid} {075437} (\bibinfo {year}
  {2009})}\BibitemShut {NoStop}%
\bibitem [{\citenamefont {Mendoza-Coto}\ \emph {et~al.}(2015)\citenamefont
  {Mendoza-Coto}, \citenamefont {Stariolo},\ and\ \citenamefont
  {Nicolao}}]{MeStNi2015}%
  \BibitemOpen
  \bibfield  {author} {\bibinfo {author} {\bibfnamefont {A.}~\bibnamefont
  {Mendoza-Coto}}, \bibinfo {author} {\bibfnamefont {D.~A.}\ \bibnamefont
  {Stariolo}}, \ and\ \bibinfo {author} {\bibfnamefont {L.}~\bibnamefont
  {Nicolao}},\ }\href {\doibase 10.1103/PhysRevLett.114.116101} {\bibfield
  {journal} {\bibinfo  {journal} {Phys. Rev. Lett.}\ }\textbf {\bibinfo
  {volume} {114}},\ \bibinfo {pages} {116101} (\bibinfo {year}
  {2015})}\BibitemShut {NoStop}%
\bibitem [{\citenamefont {Mendoza-Coto}\ \emph {et~al.}(2016)\citenamefont
  {Mendoza-Coto}, \citenamefont {Stariolo},\ and\ \citenamefont
  {Nicolao}}]{MeStNi2016}%
  \BibitemOpen
  \bibfield  {author} {\bibinfo {author} {\bibfnamefont {A.}~\bibnamefont
  {Mendoza-Coto}}, \bibinfo {author} {\bibfnamefont {D.~A.}\ \bibnamefont
  {Stariolo}}, \ and\ \bibinfo {author} {\bibfnamefont {L.}~\bibnamefont
  {Nicolao}},\ }\href {\doibase 10.1103/PhysRevLett.117.239602} {\bibfield
  {journal} {\bibinfo  {journal} {Phys. Rev. Lett.}\ }\textbf {\bibinfo
  {volume} {117}},\ \bibinfo {pages} {239602} (\bibinfo {year}
  {2016})}\BibitemShut {NoStop}%
\bibitem [{\citenamefont {Nicolao}\ \emph {et~al.}(2016)\citenamefont
  {Nicolao}, \citenamefont {Mendoza-Coto},\ and\ \citenamefont
  {Stariolo}}]{NiMeSt2016}%
  \BibitemOpen
  \bibfield  {author} {\bibinfo {author} {\bibfnamefont {L.}~\bibnamefont
  {Nicolao}}, \bibinfo {author} {\bibfnamefont {A.}~\bibnamefont
  {Mendoza-Coto}}, \ and\ \bibinfo {author} {\bibfnamefont {D.~A.}\
  \bibnamefont {Stariolo}},\ }\href
  {http://stacks.iop.org/1742-6596/686/i=1/a=012005} {\bibfield  {journal}
  {\bibinfo  {journal} {Journal of Physics: Conference Series}\ }\textbf
  {\bibinfo {volume} {686}},\ \bibinfo {pages} {012005} (\bibinfo {year}
  {2016})}\BibitemShut {NoStop}%
\bibitem [{\citenamefont {Lawler}\ \emph {et~al.}(2006)\citenamefont {Lawler},
  \citenamefont {Barci}, \citenamefont {Fern\'andez}, \citenamefont {Fradkin},\
  and\ \citenamefont {Oxman}}]{Lawler2006}%
  \BibitemOpen
  \bibfield  {author} {\bibinfo {author} {\bibfnamefont {M.~J.}\ \bibnamefont
  {Lawler}}, \bibinfo {author} {\bibfnamefont {D.~G.}\ \bibnamefont {Barci}},
  \bibinfo {author} {\bibfnamefont {V.}~\bibnamefont {Fern\'andez}}, \bibinfo
  {author} {\bibfnamefont {E.}~\bibnamefont {Fradkin}}, \ and\ \bibinfo
  {author} {\bibfnamefont {L.}~\bibnamefont {Oxman}},\ }\href {\doibase
  10.1103/PhysRevB.73.085101} {\bibfield  {journal} {\bibinfo  {journal} {Phys.
  Rev. B}\ }\textbf {\bibinfo {volume} {73}},\ \bibinfo {pages} {085101}
  (\bibinfo {year} {2006})}\BibitemShut {NoStop}%
\bibitem [{\citenamefont {Metlitski}\ and\ \citenamefont
  {Sachdev}(2010)}]{Metlitski2010}%
  \BibitemOpen
  \bibfield  {author} {\bibinfo {author} {\bibfnamefont {M.~A.}\ \bibnamefont
  {Metlitski}}\ and\ \bibinfo {author} {\bibfnamefont {S.}~\bibnamefont
  {Sachdev}},\ }\href {\doibase 10.1103/PhysRevB.82.075127} {\bibfield
  {journal} {\bibinfo  {journal} {Phys. Rev. B}\ }\textbf {\bibinfo {volume}
  {82}},\ \bibinfo {pages} {075127} (\bibinfo {year} {2010})}\BibitemShut
  {NoStop}%
\bibitem [{\citenamefont {Emery}\ and\ \citenamefont
  {Kivelson}(1993)}]{EmKi1993}%
  \BibitemOpen
  \bibfield  {author} {\bibinfo {author} {\bibfnamefont {V.}~\bibnamefont
  {Emery}}\ and\ \bibinfo {author} {\bibfnamefont {S.}~\bibnamefont
  {Kivelson}},\ }\href {\doibase 10.1016/0921-4534(93)90581-A} {\bibfield
  {journal} {\bibinfo  {journal} {Physica C: Superconductivity}\ }\textbf
  {\bibinfo {volume} {209}},\ \bibinfo {pages} {597 } (\bibinfo {year}
  {1993})}\BibitemShut {NoStop}%
\bibitem [{\citenamefont {Seul}\ and\ \citenamefont
  {Andelman}(1995)}]{SeAn1995}%
  \BibitemOpen
  \bibfield  {author} {\bibinfo {author} {\bibfnamefont {M.}~\bibnamefont
  {Seul}}\ and\ \bibinfo {author} {\bibfnamefont {D.}~\bibnamefont
  {Andelman}},\ }\href@noop {} {\bibfield  {journal} {\bibinfo  {journal}
  {Science}\ }\textbf {\bibinfo {volume} {267}},\ \bibinfo {pages} {476}
  (\bibinfo {year} {1995})}\BibitemShut {NoStop}%
\bibitem [{\citenamefont {Chayes}\ \emph {et~al.}()\citenamefont {Chayes},
  \citenamefont {Emery}, \citenamefont {Kivelson}, \citenamefont {Nussinov},\
  and\ \citenamefont {Tarjus}}]{Chayes1996}%
  \BibitemOpen
  \bibfield  {author} {\bibinfo {author} {\bibfnamefont {L.}~\bibnamefont
  {Chayes}}, \bibinfo {author} {\bibfnamefont {V.}~\bibnamefont {Emery}},
  \bibinfo {author} {\bibfnamefont {S.}~\bibnamefont {Kivelson}}, \bibinfo
  {author} {\bibfnamefont {Z.}~\bibnamefont {Nussinov}}, \ and\ \bibinfo
  {author} {\bibfnamefont {G.}~\bibnamefont {Tarjus}},\ }\href@noop {}
  {\bibfield  {journal} {\bibinfo  {journal} {Physica A}\ }\textbf {\bibinfo
  {volume} {225}},\ \bibinfo {pages} {129}}\BibitemShut {NoStop}%
\bibitem [{\citenamefont {Barci}\ \emph {et~al.}(2002)\citenamefont {Barci},
  \citenamefont {Fradkin}, \citenamefont {Kivelson},\ and\ \citenamefont
  {Oganesyan}}]{BaFrKiOg2002}%
  \BibitemOpen
  \bibfield  {author} {\bibinfo {author} {\bibfnamefont {D.~G.}\ \bibnamefont
  {Barci}}, \bibinfo {author} {\bibfnamefont {E.}~\bibnamefont {Fradkin}},
  \bibinfo {author} {\bibfnamefont {S.~A.}\ \bibnamefont {Kivelson}}, \ and\
  \bibinfo {author} {\bibfnamefont {V.}~\bibnamefont {Oganesyan}},\ }\href
  {\doibase 10.1103/PhysRevB.65.245319} {\bibfield  {journal} {\bibinfo
  {journal} {Phys. Rev. B}\ }\textbf {\bibinfo {volume} {65}},\ \bibinfo
  {pages} {245319} (\bibinfo {year} {2002})}\BibitemShut {NoStop}%
\bibitem [{\citenamefont {Bruun}\ and\ \citenamefont
  {Nelson}(2014)}]{BrNe2014}%
  \BibitemOpen
  \bibfield  {author} {\bibinfo {author} {\bibfnamefont {G.~M.}\ \bibnamefont
  {Bruun}}\ and\ \bibinfo {author} {\bibfnamefont {D.~R.}\ \bibnamefont
  {Nelson}},\ }\href {\doibase 10.1103/PhysRevB.89.094112} {\bibfield
  {journal} {\bibinfo  {journal} {Phys. Rev. B}\ }\textbf {\bibinfo {volume}
  {89}},\ \bibinfo {pages} {094112} (\bibinfo {year} {2014})}\BibitemShut
  {NoStop}%
\bibitem [{\citenamefont {Brazovskii}(1975)}]{Br1975}%
  \BibitemOpen
  \bibfield  {author} {\bibinfo {author} {\bibfnamefont {S.~A.}\ \bibnamefont
  {Brazovskii}},\ }\href@noop {} {\bibfield  {journal} {\bibinfo  {journal}
  {Sov. Phys. JETP}\ }\textbf {\bibinfo {volume} {41}},\ \bibinfo {pages} {85}
  (\bibinfo {year} {1975})}\BibitemShut {NoStop}%
\bibitem [{\citenamefont {Barci}\ and\ \citenamefont
  {Stariolo}(2007)}]{BaSt2007}%
  \BibitemOpen
  \bibfield  {author} {\bibinfo {author} {\bibfnamefont {D.~G.}\ \bibnamefont
  {Barci}}\ and\ \bibinfo {author} {\bibfnamefont {D.~A.}\ \bibnamefont
  {Stariolo}},\ }\href {\doibase 10.1103/PhysRevLett.98.200604} {\bibfield
  {journal} {\bibinfo  {journal} {Phys. Rev. Lett.}\ }\textbf {\bibinfo
  {volume} {98}},\ \bibinfo {eid} {200604} (\bibinfo {year}
  {2007})}\BibitemShut {NoStop}%
\bibitem [{\citenamefont {de~Gennes}\ and\ \citenamefont
  {Prost}(1998)}]{deGPr1998}%
  \BibitemOpen
  \bibfield  {author} {\bibinfo {author} {\bibfnamefont {P.~G.}\ \bibnamefont
  {de~Gennes}}\ and\ \bibinfo {author} {\bibfnamefont {J.}~\bibnamefont
  {Prost}},\ }\href@noop {} {\emph {\bibinfo {title} {The {P}hysics of {L}iquid
  {C}rystals}}},\ edited by\ \bibinfo {editor} {\bibfnamefont {O.~U.}\
  \bibnamefont {Press}}\ (\bibinfo  {publisher} {Oxford University Press},\
  \bibinfo {address} {Oxford},\ \bibinfo {year} {1998})\BibitemShut {NoStop}%
\bibitem [{\citenamefont {Chaikin}\ and\ \citenamefont
  {Lubensky}(1995)}]{ChLu1995}%
  \BibitemOpen
  \bibfield  {author} {\bibinfo {author} {\bibfnamefont {P.~M.}\ \bibnamefont
  {Chaikin}}\ and\ \bibinfo {author} {\bibfnamefont {T.~C.}\ \bibnamefont
  {Lubensky}},\ }\href@noop {} {\emph {\bibinfo {title} {{Principles of
  Condensed Matter Physics}}}}\ (\bibinfo  {publisher} {Cambridge University
  Press},\ \bibinfo {address} {Cambridge, UK},\ \bibinfo {year}
  {1995})\BibitemShut {NoStop}%
\bibitem [{\citenamefont {Cvetkovic}\ \emph {et~al.}(2006)\citenamefont
  {Cvetkovic}, \citenamefont {Nussinov},\ and\ \citenamefont
  {Zaanen}}]{Nussinov2006}%
  \BibitemOpen
  \bibfield  {author} {\bibinfo {author} {\bibfnamefont {V.}~\bibnamefont
  {Cvetkovic}}, \bibinfo {author} {\bibfnamefont {Z.}~\bibnamefont {Nussinov}},
  \ and\ \bibinfo {author} {\bibfnamefont {J.}~\bibnamefont {Zaanen}},\
  }\href@noop {} {\bibfield  {journal} {\bibinfo  {journal} {Philosophical
  Magazine}\ }\textbf {\bibinfo {volume} {86}},\ \bibinfo {pages} {2995}
  (\bibinfo {year} {2006})}\BibitemShut {NoStop}%
\bibitem [{\citenamefont {Golubovi\ifmmode~\acute{c}\else \'{c}\fi{}}\ \emph
  {et~al.}(2000)\citenamefont {Golubovi\ifmmode~\acute{c}\else \'{c}\fi{}},
  \citenamefont {Lubensky},\ and\ \citenamefont {O'Hern}}]{Lubensky2000}%
  \BibitemOpen
  \bibfield  {author} {\bibinfo {author} {\bibfnamefont {L.}~\bibnamefont
  {Golubovi\ifmmode~\acute{c}\else \'{c}\fi{}}}, \bibinfo {author}
  {\bibfnamefont {T.~C.}\ \bibnamefont {Lubensky}}, \ and\ \bibinfo {author}
  {\bibfnamefont {C.~S.}\ \bibnamefont {O'Hern}},\ }\href {\doibase
  10.1103/PhysRevE.62.1069} {\bibfield  {journal} {\bibinfo  {journal} {Phys.
  Rev. E}\ }\textbf {\bibinfo {volume} {62}},\ \bibinfo {pages} {1069}
  (\bibinfo {year} {2000})}\BibitemShut {NoStop}%
\bibitem [{\citenamefont {Barci}\ \emph {et~al.}(2013)\citenamefont {Barci},
  \citenamefont {Mendoza-Coto},\ and\ \citenamefont {Stariolo}}]{BaMeSt2013}%
  \BibitemOpen
  \bibfield  {author} {\bibinfo {author} {\bibfnamefont {D.~G.}\ \bibnamefont
  {Barci}}, \bibinfo {author} {\bibfnamefont {A.}~\bibnamefont {Mendoza-Coto}},
  \ and\ \bibinfo {author} {\bibfnamefont {D.~A.}\ \bibnamefont {Stariolo}},\
  }\href {\doibase 10.1103/PhysRevE.88.062140} {\bibfield  {journal} {\bibinfo
  {journal} {Phys. Rev. E}\ }\textbf {\bibinfo {volume} {88}},\ \bibinfo
  {pages} {062140} (\bibinfo {year} {2013})}\BibitemShut {NoStop}%
\bibitem [{\citenamefont {Maier}\ and\ \citenamefont
  {Schwabl}(2004)}]{MaSc2004}%
  \BibitemOpen
  \bibfield  {author} {\bibinfo {author} {\bibfnamefont {P.~G.}\ \bibnamefont
  {Maier}}\ and\ \bibinfo {author} {\bibfnamefont {F.}~\bibnamefont
  {Schwabl}},\ }\href {\doibase 10.1103/PhysRevB.70.134430} {\bibfield
  {journal} {\bibinfo  {journal} {Phys. Rev. B}\ }\textbf {\bibinfo {volume}
  {70}},\ \bibinfo {pages} {134430} (\bibinfo {year} {2004})}\BibitemShut
  {NoStop}%
\bibitem [{\citenamefont {Br\'ezin}\ and\ \citenamefont
  {Zinn-Justin}(1976)}]{Brzi1976}%
  \BibitemOpen
  \bibfield  {author} {\bibinfo {author} {\bibfnamefont {E.}~\bibnamefont
  {Br\'ezin}}\ and\ \bibinfo {author} {\bibfnamefont {J.}~\bibnamefont
  {Zinn-Justin}},\ }\href {\doibase 10.1103/PhysRevB.14.3110} {\bibfield
  {journal} {\bibinfo  {journal} {Phys. Rev. B}\ }\textbf {\bibinfo {volume}
  {14}},\ \bibinfo {pages} {3110} (\bibinfo {year} {1976})}\BibitemShut
  {NoStop}%
\bibitem [{\citenamefont {Br\'ezin}\ \emph {et~al.}(1976)\citenamefont
  {Br\'ezin}, \citenamefont {Le~Guillou},\ and\ \citenamefont
  {Zinn-Justin}}]{BrLe1976}%
  \BibitemOpen
  \bibfield  {author} {\bibinfo {author} {\bibfnamefont {E.}~\bibnamefont
  {Br\'ezin}}, \bibinfo {author} {\bibfnamefont {J.~C.}\ \bibnamefont
  {Le~Guillou}}, \ and\ \bibinfo {author} {\bibfnamefont {J.}~\bibnamefont
  {Zinn-Justin}},\ }\href@noop {} {\emph {\bibinfo {title} {Phase Transitions
  and Critical Phenomena}}},\ Vol.~\bibinfo {volume} {6}\ (\bibinfo
  {publisher} {Academic Press, London},\ \bibinfo {year} {1976})\BibitemShut
  {NoStop}%
\bibitem [{\citenamefont {Amit}(1978)}]{Amit1978}%
  \BibitemOpen
  \bibfield  {author} {\bibinfo {author} {\bibfnamefont {D.~J.}\ \bibnamefont
  {Amit}},\ }\href@noop {} {\emph {\bibinfo {title} {{Field Theory, the
  Renormalization Group, and Critical Phenomena}}}}\ (\bibinfo  {publisher}
  {McGraw-Hill International Book Company},\ \bibinfo {address} {New York},\
  \bibinfo {year} {1978})\BibitemShut {NoStop}%
\bibitem [{\citenamefont {Dutta}\ and\ \citenamefont
  {Bhattacharjee}(2001)}]{DuBh2001}%
  \BibitemOpen
  \bibfield  {author} {\bibinfo {author} {\bibfnamefont {A.}~\bibnamefont
  {Dutta}}\ and\ \bibinfo {author} {\bibfnamefont {J.~K.}\ \bibnamefont
  {Bhattacharjee}},\ }\href {\doibase 10.1103/PhysRevB.64.184106} {\bibfield
  {journal} {\bibinfo  {journal} {Phys. Rev. B}\ }\textbf {\bibinfo {volume}
  {64}},\ \bibinfo {pages} {184106} (\bibinfo {year} {2001})}\BibitemShut
  {NoStop}%
\bibitem [{\citenamefont {Hooley}\ \emph {et~al.}(2014)\citenamefont {Hooley},
  \citenamefont {Carr}, \citenamefont {Fellows},\ and\ \citenamefont
  {Schmalian}}]{HoCa2014}%
  \BibitemOpen
  \bibfield  {author} {\bibinfo {author} {\bibfnamefont {C.}~\bibnamefont
  {Hooley}}, \bibinfo {author} {\bibfnamefont {S.}~\bibnamefont {Carr}},
  \bibinfo {author} {\bibfnamefont {J.}~\bibnamefont {Fellows}}, \ and\
  \bibinfo {author} {\bibfnamefont {J.}~\bibnamefont {Schmalian}},\ }\href
  {\doibase http://dx.doi.org/10.7566/JPSCP.3.016018} {\bibfield  {journal}
  {\bibinfo  {journal} {Proceedings of the International Conference on Strongly
  Correlated Electron Systems (SCES2013)}\ }\textbf {\bibinfo {volume}
  {016018}} (\bibinfo {year} {2014}),\
  http://dx.doi.org/10.7566/JPSCP.3.016018}\BibitemShut {NoStop}%
\bibitem [{\citenamefont {Campostrini}\ \emph {et~al.}(2001)\citenamefont
  {Campostrini}, \citenamefont {Hasenbusch}, \citenamefont {Pelissetto},
  \citenamefont {Rossi},\ and\ \citenamefont {Vicari}}]{CaHa2001}%
  \BibitemOpen
  \bibfield  {author} {\bibinfo {author} {\bibfnamefont {M.}~\bibnamefont
  {Campostrini}}, \bibinfo {author} {\bibfnamefont {M.}~\bibnamefont
  {Hasenbusch}}, \bibinfo {author} {\bibfnamefont {A.}~\bibnamefont
  {Pelissetto}}, \bibinfo {author} {\bibfnamefont {P.}~\bibnamefont {Rossi}}, \
  and\ \bibinfo {author} {\bibfnamefont {E.}~\bibnamefont {Vicari}},\ }\href
  {\doibase 10.1103/PhysRevB.63.214503} {\bibfield  {journal} {\bibinfo
  {journal} {Phys. Rev. B}\ }\textbf {\bibinfo {volume} {63}},\ \bibinfo
  {pages} {214503} (\bibinfo {year} {2001})}\BibitemShut {NoStop}%
\bibitem [{\citenamefont {Langfeld}(2013)}]{Langf2013}%
  \BibitemOpen
  \bibfield  {author} {\bibinfo {author} {\bibfnamefont {K.}~\bibnamefont
  {Langfeld}},\ }\href {\doibase 10.1103/PhysRevD.87.114504} {\bibfield
  {journal} {\bibinfo  {journal} {Phys. Rev. D}\ }\textbf {\bibinfo {volume}
  {87}},\ \bibinfo {pages} {114504} (\bibinfo {year} {2013})}\BibitemShut
  {NoStop}%
\bibitem [{\citenamefont {McMillan}(1971)}]{McMi1971}%
  \BibitemOpen
  \bibfield  {author} {\bibinfo {author} {\bibfnamefont {W.~L.}\ \bibnamefont
  {McMillan}},\ }\href {\doibase 10.1103/PhysRevA.4.1238} {\bibfield  {journal}
  {\bibinfo  {journal} {Phys. Rev. A}\ }\textbf {\bibinfo {volume} {4}},\
  \bibinfo {pages} {1238} (\bibinfo {year} {1971})}\BibitemShut {NoStop}%
\bibitem [{\citenamefont {Sun}\ \emph {et~al.}(2008)\citenamefont {Sun},
  \citenamefont {Fregoso}, \citenamefont {Lawler},\ and\ \citenamefont
  {Fradkin}}]{Lawler2008}%
  \BibitemOpen
  \bibfield  {author} {\bibinfo {author} {\bibfnamefont {K.}~\bibnamefont
  {Sun}}, \bibinfo {author} {\bibfnamefont {B.~M.}\ \bibnamefont {Fregoso}},
  \bibinfo {author} {\bibfnamefont {M.~J.}\ \bibnamefont {Lawler}}, \ and\
  \bibinfo {author} {\bibfnamefont {E.}~\bibnamefont {Fradkin}},\ }\href
  {\doibase 10.1103/PhysRevB.78.085124} {\bibfield  {journal} {\bibinfo
  {journal} {Phys. Rev. B}\ }\textbf {\bibinfo {volume} {78}},\ \bibinfo
  {pages} {085124} (\bibinfo {year} {2008})}\BibitemShut {NoStop}%
\bibitem [{\citenamefont {Sun}\ \emph {et~al.}(2009)\citenamefont {Sun},
  \citenamefont {Fregoso}, \citenamefont {Lawler},\ and\ \citenamefont
  {Fradkin}}]{Lawler-Erratum2009}%
  \BibitemOpen
  \bibfield  {author} {\bibinfo {author} {\bibfnamefont {K.}~\bibnamefont
  {Sun}}, \bibinfo {author} {\bibfnamefont {B.~M.}\ \bibnamefont {Fregoso}},
  \bibinfo {author} {\bibfnamefont {M.~J.}\ \bibnamefont {Lawler}}, \ and\
  \bibinfo {author} {\bibfnamefont {E.}~\bibnamefont {Fradkin}},\ }\href
  {\doibase 10.1103/PhysRevB.80.039901} {\bibfield  {journal} {\bibinfo
  {journal} {Phys. Rev. B}\ }\textbf {\bibinfo {volume} {80}},\ \bibinfo
  {pages} {039901} (\bibinfo {year} {2009})}\BibitemShut {NoStop}%
\bibitem [{\citenamefont {Barci}\ and\ \citenamefont
  {Fradkin}(2011)}]{BarciFradkin2011}%
  \BibitemOpen
  \bibfield  {author} {\bibinfo {author} {\bibfnamefont {D.~G.}\ \bibnamefont
  {Barci}}\ and\ \bibinfo {author} {\bibfnamefont {E.}~\bibnamefont
  {Fradkin}},\ }\href {\doibase 10.1103/PhysRevB.83.100509} {\bibfield
  {journal} {\bibinfo  {journal} {Phys. Rev. B}\ }\textbf {\bibinfo {volume}
  {83}},\ \bibinfo {pages} {100509} (\bibinfo {year} {2011})}\BibitemShut
  {NoStop}%
\bibitem [{\citenamefont {McMillan}(1972)}]{McMi1972}%
  \BibitemOpen
  \bibfield  {author} {\bibinfo {author} {\bibfnamefont {W.~L.}\ \bibnamefont
  {McMillan}},\ }\href {\doibase 10.1103/PhysRevA.6.936} {\bibfield  {journal}
  {\bibinfo  {journal} {Phys. Rev. A}\ }\textbf {\bibinfo {volume} {6}},\
  \bibinfo {pages} {936} (\bibinfo {year} {1972})}\BibitemShut {NoStop}%
\bibitem [{\citenamefont {McMillan}(1973)}]{McMi1973}%
  \BibitemOpen
  \bibfield  {author} {\bibinfo {author} {\bibfnamefont {W.~L.}\ \bibnamefont
  {McMillan}},\ }\href {\doibase 10.1103/PhysRevA.7.1673} {\bibfield  {journal}
  {\bibinfo  {journal} {Phys. Rev. A}\ }\textbf {\bibinfo {volume} {7}},\
  \bibinfo {pages} {1673} (\bibinfo {year} {1973})}\BibitemShut {NoStop}%
\end{thebibliography}
%
%
%
\end{document}